\documentclass[10pt,notitlepage]{article}
\usepackage[utf8x]{inputenc}
\usepackage[a4paper, margin=2cm]{geometry}
\usepackage{authblk}

\usepackage{amsfonts}
\usepackage{amsmath}
\usepackage{amsthm}
\usepackage{amssymb}
\usepackage{array}
\usepackage{dcolumn}
\usepackage[svgnames]{xcolor}
\usepackage[hidelinks]{hyperref}
\hypersetup{
    colorlinks=true,
    linkcolor=NavyBlue,
    filecolor=NavyBlue,
    urlcolor=NavyBlue,
    citecolor=NavyBlue,
    }

\usepackage{graphics}
\usepackage{graphicx}
\usepackage{caption}
\usepackage{subcaption}
\usepackage{adjustbox}
\usepackage{gensymb}
\usepackage{breqn}
\usepackage{braket}
\usepackage{enumitem}
\usepackage{makecell}
\usepackage{ragged2e}
\usepackage[sort&compress,numbers]{natbib}

\usepackage{textpos}

\def\gev{\, {\rm GeV}}
\def\tev{\, {\rm TeV}}
\def\mev{\, {\rm MeV}}

\def\be{\begin{equation}}
\def\ee{\end{equation}}
\def\bea{\begin{eqnarray}}
\def\eea{\end{eqnarray}}

\newcommand\snowmass{
\begin{center}
  \rule[-0.2in]{\hsize}{0.01in}\\
  \rule{\hsize}{0.01in}\\
  \vskip 0.1in
  Submitted to the Proceedings of the US Community Study\\ 
  on the Future of Particle Physics (Snowmass 2021)\\
  \rule{\hsize}{0.01in}\\
  \rule[+0.2in]{\hsize}{0.01in}\\[-2em]
\end{center}
}

\usepackage[firstpage=true]{background}
\backgroundsetup{contents={\parbox{6.5in}{\snowmass}}, scale=1,placement=top,opacity=1,color=black,position={3.25in,1.2in}}

\usepackage{fancyhdr}
\fancypagestyle{plain}{%
  \fancyhf{}%
  \fancyhead[C]{}
  \fancyfoot[C]{\thepage}
}

\fancypagestyle{empty}{%
  \fancyfoot[C]{\thepage}
}
\title{\large{\textbf{ $U(1)_{T3R}$ Extension of Standard Model: A Sub-GeV Dark Matter Model}}}
\date{}

\usepackage{authblk}

\author[1]{Bhaskar Dutta\footnote{dutta@physics.tamu.edu}}
\author[1,2]{Sumit Ghosh\footnote{ghosh@kias.re.kr/ghosh@tamu.edu}}
\author[3]{Jason Kumar\footnote{jkumar@hawaii.edu}}
 
\affil[1]{Mitchell Institute for Fundamental Physics and Astronomy, Department of Physics  and Astronomy, Texas A$\&$M University,
College Station, Texas 77843, USA}
\affil[2]{School of Physics,  Korea Institute for Advanced Study,
Seoul 02455, Korea}
\affil[3]{Department of Physics, University of Hawaii, Honolulu, Hawaii 96822, USA }

\begin{document}

\maketitle

\begin{abstract}

We present a model based on a $U(1)_{T3R}$ extension of the Standard Model. The model addresses the mass hierarchy between the third generation  and the first two generation fermions. $U(1)_{T3R}$ is spontaneously broken at $\sim 1-10~\gev$. The model contains a sub-GeV dark matter candidate and two sub-GeV light scalar and vector mediators. The model explains the thermal dark matter abundance, measurements of the muon $g-2$ and $R_{K^{(\ast)}}$ anomalies. The model can be probed at the LHC, FASER, dark matter experiments and various beam-dump based neutrino facilities, e.g., COHERENT, CCM, MicroBooNE, SBND, ICARUS, DUNE etc.
\end{abstract}

\section{Introduction} \label{sec:intro}

The origins of dark matter~\cite{Planck:2018vyg}, tiny neutrino masses~\cite{Fukuda:1998mi,Ahmad:2002jz}, electroweak symmetry breaking scales and various anomalies, e.g., $g-2$ of the muon~\cite{aoyama:2012wk, Aoyama:2019ryr, czarnecki:2002nt, gnendiger:2013pva, Davier:2017zfy, Blum:2018mom,  Keshavarzi:2018mgv, colangelo:2018mtw, hoferichter:2019gzf, Davier:2019can, Keshavarzi:2019abf, kurz:2014wya, melnikov:2003xd, masjuan:2017tvw, Colangelo:2017fiz, hoferichter:2018kwz, gerardin:2019vio, bijnens:2019ghy, colangelo:2019uex, Blum:2019ugy, colangelo:2014qya, Campanario:2019mjh, Abi:2021gix, Bennett:2006fi, Borsanyi:2020mff, Crivellin:2020zul,Keshavarzi:2020bfy,Colangelo:2020lcg}, $R_{K^{(\ast})}$~\cite{RKstar,RKstarBelle, Aaij:2021vac}, excess observed at MiniBooNE~\cite{Aguilar-Arevalo:2018gpe, Aguilar-Arevalo:2020nvw} etc. are still not known. The scale of new physics associated with the possible explanations for these puzzles are being searched at various ongoing experiments. For example, LHC is searching for new physics at $\sim$TeV scale, various indirect and direct detection experiments are searching for new physics scales from sub-GeV (or lower) to TeV and various low energy neutrino experiments (beam-dump and reactor) experiments are probing scales even lower than sub-GeV. 

Since a wide range of new physics scales are being probed at various ongoing experiments, we propose a model  containing sub-GeV dark matter  motivated by a solution for the Standard Model (SM) fermion mass hierarchy problem. The model has light gauge and scalar particles associated with new gauge group $U(1)_{T3R}$. This model also involves TeV scale particle in order to provide an UV completion for the origin of fermion mass and its hierarchies. The model has the potential to explain the MiniBooNE, $g-2$, and  $R_{K^{(\ast)}}$ anomalies as well as to provide a the thermal dark matter candidate. The model can be probed at the direct and indirect detection experiments. Innovative search techniques can make this model detectable at the LHC and at various low energy experiments.

In many extensions of the SM, a new gauge group has been proposed which leads to  interactions between Standard Model particles and new sub-GeV particles. The introduction of new gauge interactions also require that gauge and gravitational anomalies to be cancelled.  There exists many well studied examples which include, $U(1)_{B-L}$, $U(1)_{L_i - L_j}$,  a secluded $U(1)_X$ (under which all Standard Model particles are neutral) etc~\cite{Foot:1990mn, He:1990pn, He:1991qd, Borah:2020swo, Costa:2020krs}.  Recently we studied  $U(1)_{T3R}$~\cite{Dutta:2019fxn, Dutta:2020enk, Dutta:2020jsy, Dutta:2021afo, Dutta:2022bfq},  which contains one or more full generations of right-handed Standard Model fermions which are charged (including right-handed neutrinos) under this new symmetry, with up- and down-type fermions having opposite charge.

$U(1)_{T3R}$ was studied originally in the context of left-right models (for example,~\cite{Pati:1974yy, Mohapatra:1974gc, Senjanovic:1975rk}).  In this extension, the Standard Model Higgs has $U(1)_{T3R}$ charge which connects  the symmetry-breaking scale of $U(1)_{T3R}$ to the electroweak-scale.  Recent interests are focused on scenarios where the $U(1)_{T3R}$ is broken by a dark Higgs, which provides a new independent scale  decoupled from the EW-scale.  Only the right-handed fermions are charged under $U(1)_{T3R}$ and hence the fermion masses are also protected by this symmetry, and are thus proportional to this new symmetry-breaking scale.  In the recent works, a  new symmetry-breaking scale $V \sim {\cal O}(10)\gev$ is considered where only the first- or second-generation fermions are charged under $U(1)_{T3R}$~\cite{Dutta:2019fxn}.  In this scenario, the Yukawa couplings of the low energy effective field theory (defined below the electroweak scale) can be ${\cal O}(0.1-0.01)$ thus providing an explanation for the Standard Model fermion mass hierarchy problem.  Additionally,  the mass scale of dark sector particles which are only charged under $U(1)_{T3R}$ is naturally set by this new symmetry-breaking scale, while being singlets under Standard Model gauge groups.  This scenario  motivates the appearance of new sub-GeV particles from the solution of the fermion mass hierarchy problem.

There are a few key theoretical features which are worth noting.\begin{itemize}\item{ $U(1)_{T3R}$ protects the masses of Standard Model fermions where  the dark Higgs whose vev breaks $U(1)_{T3R}$ must couple to Standard Model fermions, in addition to the dark gauge boson.  This scenario thus  contains two types of light mediators which interact with the SM fermions, unlike most other examples of new gauge groups investigated with sub-GeV mediator.}\item{Since the dark photon has chiral couplings to Standard Model fermions, the longitudinal polarization does not decouple. Since  this mode has its origin as a Goldstone boson, The couplings of the dark photon arethus related to those of dark Higgs  } \item{Since the SM fermions which couple to $U(1)_{T3R}$ have masses which are not much smaller than the symmetry-breaking scale, the Yukawa couplings of the low-energy effective field theory are not very small.  Thus, the dark Higgs (and, necessarily, the dark photon) must have relatively large couplings to the Standard Model particles.}\end{itemize}

The fact that these couplings are actually reasonably large creates a window of opportunity for  various experiments. Since, the $U(1)_{T3R}$ scenario necessarily contains two mediators, a scalar and a vector, the muon magnetic moment gets opposite contributions  since the dark photon has an axial coupling.  Further, since the coupling of the dark photon is tied to that of the dark scalar, one generally finds regions of parameter space in which the scalar and vector contributions cancel giving rise to larger couplings compared to the single mediator models.  One  interesting way to constrain this model is with displaced detectors which are nevertheless close enough to interaction point that the dark mediators can reach the detector before decaying.  The light scalar mediator of the model also can be probed at the LHC utilizing heavy top partner which is present in the model as a part of UV completion . The production of the light particle emerges from the decay of the heavier particle which allows it to possess large transverse momentum.  

The light mediators of this model also can be probed at various beam-dump based neutrino experiments, COHERENT~\cite{COHERENT:2017ipa,COHERENT:2018imc,COHERENT:2018gft,COHERENT:2019kwz,COHERENT:2020iec, Akimov:2017ade, Akimov:2018vzs, Akimov:2018ghi, Akimov:2019xdj, Akimov:2020pdx}, CCM~\cite{CCM1, CCM2}, MicroBooNE~\cite{MicroBooNE:2016pwy, MicroBooNE:2019lta, MicroBooNE:2021nxr, MicroBooNE:2021rmx, MicroBooNE:2021sne, MicroBooNE:2021jwr}, ICARUS~\cite{MicroBooNE:2015bmn}, SBND~\cite{SBND:2020eho, SBND:2020scp}, DUNE~\cite{Strait:2015aku, Habig:2015rop, Abi:2020wmh, Abi:2020evt, Abi:2020oxb, Abi:2020loh} etc. In these facilities, the quark couplings will allow the mediators to be produced from the charged and neutral pion and kaon decays at the detector the mediators can be detected from the their visible, invisible decays products and inverse -Primakoff type interactions.

The rest of the paper is organized as follows: In Sec.~\ref{sec:model}, we describe the model details including mass generations and interaction terms. In Sec.~\ref{sec:UV}, we discuss possible UV completion of the model. We discuss constraints in Sec.~\ref{sec:constraints}. The direct detection prospect is described in Sec.~\ref{sec:direct detection}. In Sec.~\ref{sec:relic density}, we discuss how  the correct relic density can be obtained in this model. In Sec.~\ref{sec:flavor physics}, how the flavor physics anomalies related to B-physics can be accommodated in this model. In Sec.~\ref{sec:detection}, we discuss detection prospects of this model in various upcoming/ongoing experiments. We conclude in Sec.~\ref{sec:conclusion}.

\section{$U(1)_{T3R}$ Model} \label{sec:model}

The details of this scenario are explained in Refs.~\cite{Dutta:2019fxn,Dutta:2020jsy}, but we will briefly review the salient points. We extend the gauge symmetry of SM by an extra abelian gauge group, $U(1)_{}T3R$. This gauge group was first introduced in the context of left-right symmetric model~\cite{Pati:1974yy, Mohapatra:1974gc, Senjanovic:1975rk}. Here we utilize this to explore low energy physics. Therefore the complete low energy gauge symmetry of our model is $SU(3)_C$$\times$$SU(2)_L$$\times$$U(1)_Y$$\times$$U(1)_{T_{3R}}$. The new gauge group is not connected to the electric charge. Only the right handed SM fermions are charged under this new gauge group including a new right handed neutrino. All other Sm fields have theier usual charges under SM gauge groups. In addition to this, we introduce three more fields, one complex scalar singlet $\phi$, and a left and right-handed fermion pair $\eta_L$ and $\eta_R$. They are only charged under $U(1)_{T3R}$.

In order to ensure that all gauge and gravitational anomalies are cancelled, we will assume that one right-handed up-type quark, down-type quark, charged lepton and neutrino are charged under $U(1)_{T3R}$ with $Q= \pm 2$, and with up-type and down-type fermions having opposite sign.  Note that, although these Standard Model fermions constitute a full generation, they need not all be in the same generation.  It is technically natural for the charged lepton and either the up-type or down-type quark charged under $U(1)_{T3R}$ to be a mass eigenstate~\cite{Batell:2017kty}.  For simplicity, we will assume that all fermions charged under $U(1)_{T3R}$ are mass eigenstates. The detail charge distribution is shown in Table.~\ref{tab:charges}. We consider the case in which the Standard Model fermions which are charged under $U(1)_{T3R}$ are $u$, $d$ and  $\mu$.  This case is interesting because it avoids tight constraints which arise from atomic parity violation experiments~\cite{Diener:2011jt} and cosmological observables~\cite{Dutta:2020jsy} (if the dark photon couples to electrons) as well as constraints on the anomalous kaon decay (if the dark photon couples to second-generation quarks).

\begin{table}[h]

\captionsetup{justification   = RaggedRight,
             labelfont = bf}
\caption{ \label{tab:charges} The  charges of the fields 
under $U(1)_{T3R}$. For the fermionic fields, we list the charges of the left-handed component of the Weyl spinor.}             
\centering
\begin{tabular}{ llllllll }
\hline\hline
field&$u_R$&$d_R$& $\mu_R$&$\nu_R$&$\eta_L$&$\eta_R$&$\phi$ \\ \hline
&&&&&&&\\
$q_{T3R}$&-2&2&2&-2&1&-1&2\\
\hline\hline
\end{tabular}
\end{table}

$U(1)_{T3R}$ will be broken to a parity by the condensation of the complex scalar field $\phi$ with charge $Q_\phi = 2$. We consider the case in which $\phi$ has a quartic potential which can be written as \bea V_\phi &=& \mu_\phi^2 \phi \phi^*  + \lambda_\phi (\phi \phi^*)^2 , \eea We may then express $\phi$ as $\phi = V + (1/\sqrt{2}) (\phi' + \imath \sigma)$, where $V$ is taken to be real.  The real scalar fields $\phi'$ and $\sigma$ are the dark Higgs and the Goldstone boson, respectively. we find $V = (-\mu_\phi / 2\lambda_\phi)^{1/2}$, $m_{\phi'}^2 = -\mu_\phi^2 = 2\lambda_\phi V^2$. Note that, all the SM fields and $\phi^\prime$ are even under the parity while only $\eta_{L,R}$ are odd.

The low energy non-renormalizable  interaction Lagrangian can be written as, 
\bea \label{intlag} \mathcal{L} &=&- \frac{\lambda_u}{\Lambda} \tilde{H} \phi^* \bar{Q}_L  u_R  -  \frac{\lambda_d}{\Lambda} H \phi \bar{Q}_L d_R - \frac{\lambda_\nu}{\Lambda} \tilde{H} \phi^*\bar{L}_L  \nu_R  - \frac{\lambda_\mu}{\Lambda} H \phi \bar{L}_L \mu_R - m_D \bar{\eta}_R  \eta_L \nonumber\\  &~& - \frac{1}{2}\lambda_L \phi \bar{\eta}^c_L \eta_L - \frac{1}{2}\lambda_R \phi^* \bar{\eta}^c_R \eta_R +H.c. , \eea where $Q_L$ and $L_L$ are the left-handed SM quark and lepton doublet, respectively; and $H$ is the SM Higgs doublet; and  $\tilde{H}$$=$$i\tau_2H^*$.

A dark matter candidate naturally arises in this scenario. Dirac fermion $\eta$ which is charged only under $U(1)_{T3R}$ with charge $Q_\eta =1$ has both Dirac and Majorana mass term. If the Dirac mass, $m_D$ is very small compared to the Majorana mass, $m_M$ and we assume that $\lambda_L = \lambda_R \equiv \lambda_M$ i.e.  the Majorana masses for the left-handed and the right-handed fields are equal, with $m_M=\lambda_L V= \lambda_RV=(\lambda_M V)$, then we are left with two dark sector Majorana fermion mass eigenstates $\eta_{1,2}$, with mass $\propto V$. The physical states can be expressed as, \bea \eta_1 &=& \frac{1}{\sqrt{2}}\left( \begin{array}{c} \eta_L-\eta^c_R \\ \eta^c_L-\eta_R \end{array} \right) , ~~~~~~~~~~~~~~~~~~~~                                                                                                      \eta_2 = \frac{1}{\sqrt{2}}\left( \begin{array}{c} \eta_L+\eta^c_R \\ \eta^c_L+\eta_R \end{array} \right) ,    	\eea The masses are $m_1 = m_M-m_D $ and $m_2=m_M+m_D$ respectively and the mass splitting $\delta = 2m_D$ is very small. The small $m_D$ also makes sure that the couplings of $\phi^\prime$ to $\eta_{1,2}$ are proportional to their mass $m_{1,2}$.The lightest is  stable due to odd parity, and is a dark matter candidate.

We assume that $U(1)_{T3R}$ is broken well below the electroweak symmetry-breaking scale.  In the low energy effective field theory defined below electroweak symmetry breaking, the mass and Yukawa coupling of the fermions  charged under $U(1)_{T3R}$ arise from, \bea{\cal L} &=&-m_u\bar{u}_L u_R  -m_d\bar{d}_L d_R  -m_{\nu D}\bar{\nu}_L \nu_R -m_\mu \bar{\mu}_L \mu_R   \nonumber\\ &~& -\frac{m_u }{V\sqrt{2}}\bar{u}_L u_R \phi^\prime -\frac{m_d  }{V\sqrt{2}}\bar{d}_L d_R \phi^\prime -\frac{m_{\nu D}  }{V\sqrt{2}}\bar{\nu}_L \nu_R \phi^\prime	-\frac{m_\mu }{V\sqrt{2}}\bar{\mu}_L \mu_R \phi^\prime  \nonumber\\ &~&-\frac{1}{2} m_1 \bar{\eta}_1  {\eta}_1 -\frac{1}{2}m_2 \bar{\eta}_2  {\eta}_2 -\frac{1}{2\sqrt{2}}\frac{m_1}{V} \bar{\eta}_1  {\eta}_1 \phi^\prime-\frac{1}{2\sqrt{2}}\frac{m_2}{V} \bar{\eta}_2  {\eta}_2 \phi^\prime + H.c.  ,\eea  We thus see that if $V$ is only slightly above the mass scale of the fermions, the Yukawa coupling $\lambda_f$ need not be unnaturally small.

The neutrinos also have both Dirac mass term, $m_{\nu_D}$, which is proportional to $V$, and   Majorana mass term, which is proportional to $ V^2 / \Lambda$, where $\Lambda$ is some high-energy scale. We expect that the Majorana mass is less than $V$. The diagonalization of the neutrino mass matrix thus gives two mass eigenstates, $\nu_A$ and $\nu_S$.  We  assume small mixing between the two states such that the active neutrino $\nu_A$ is mostly $\nu_L$, with only a small mixing of $\nu_R$.

\begin{figure}[h]
\begin{subfigure}[h]{0.48\textwidth}
\includegraphics[width=.9\linewidth,height=3.2cm]{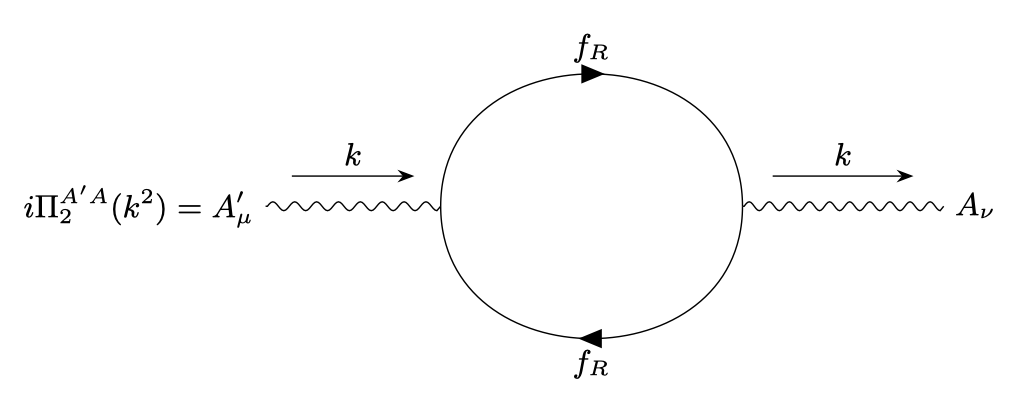}
\captionsetup{labelfont = bf}
\caption{\label{fig:loop1}}
\end{subfigure}	
\hspace{0.2cm}	
\begin{subfigure}[h]{0.48\textwidth}
\includegraphics[width=.9\linewidth,height=3.2cm]{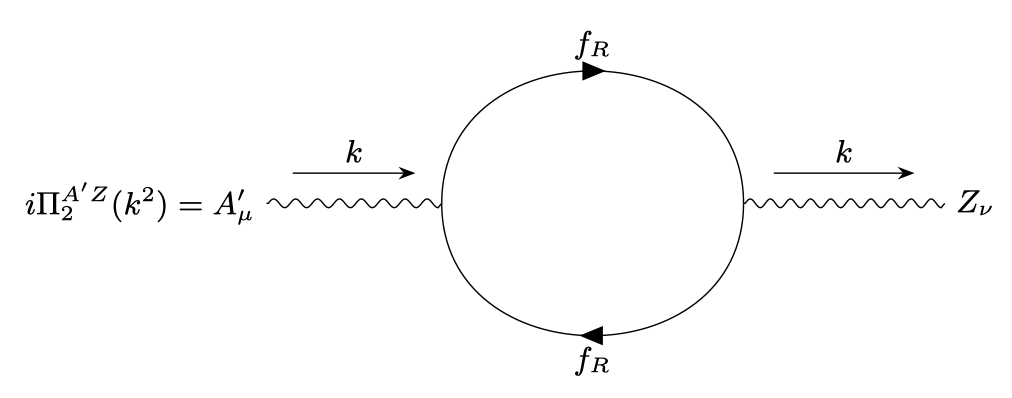}	
\captionsetup{labelfont = bf}
\caption{\label{fig:loop2}}
\end{subfigure}	

\captionsetup{justification   = RaggedRight,
             labelfont = bf}
\caption{\label{fig:loop} The one loop diagrams that lead to the kinetic mixing. } 
\end{figure}

The gauge sector of the model can be studied by defining the covariant derivative, \bea  D_{\mu}{I} = {\partial}_{\mu}{I} +i\frac{g}{2}{\tau}_a W_{\mu a}+ig^{\prime} Y B_{\mu} +i\frac{g_{T_{3R}}}{2}Q_{T_{3R}} A'_\mu. \eea where $g$, $g^{\prime}$ and $g_{T3R}$ are the coupling constants of the $SU(2)_L$, $U(1)_Y$ and $U(1)_{T3R}$ groups respectively. $W_{\mu }$, $ B_{\mu}$ and $A'_\mu$ are the gauge bosons of the $SU(2)_L$, $U(1)_Y$ and $U(1)_{T{3R}}$ gauge groups respectively. The term, $|D_{\mu}\phi|^2$ gives the dark photon mass, $m_{A^\prime}^2=2g_{T{3R}}^2 V^2$. The dark photon, $A^'$ interactions with the fermions and dark Higgs are given by, \bea \label{Aprimeint} \mathcal{L}_{\text{gauge}}&=&  \frac{m_{A'}}{ 4\sqrt{2} V}A^\prime_\mu(\bar{\eta}_1\gamma^\mu\eta_2-\bar{\eta}_2\gamma^\mu\eta_1)  +  \frac{m_{A'}^2}{V\sqrt{2}} \phi' A'_\mu A'{}^{\mu} + \frac{m_{A'}^2}{4 V^2} \phi' \phi' A'_\mu {A'}^{\mu}\nonumber\\&~& - \frac{m_{A'}}{ 2\sqrt{2}  V}j^\mu_{A^\prime}{A^\prime}_\mu . \eea The SM fermion current is defined as, $j^\mu_{A^\prime}=\sum\limits_f Q_{T_{3R}}^f \bar{f}\gamma^\mu \left( \frac{1+\gamma_5}{2} \right) f$. Also note that, the $\eta$ fields have only off-diagonal vector interaction with $A'$. In addition to these, the dark photon can couple to all the SM fermions through kinetic mixing with a coupling $\epsilon e$, where $\epsilon$ is a kinetic mixing parameter. The kinetic mixing can arise at one-loop level as shown in Fig.~\ref{fig:loop}, where the right handed fermions charged under $U(1)_{T3R}$ run inside the loop.

As we have seen, we have a scenario in which  we have two new mediators (along with a dark matter candidate) whose masses are all $\lesssim V$.  As a benchmark, we will take the symmetry-breaking scale $V = 10~\gev$.  In this case, the dark Higgs coupling to muons is $\sim m_\mu / V \sim 10^{-2}$.  We will then find that the most interesting case is $m_{A',\phi'} < 2 m_\mu$, as otherwise the mediators would decay promptly to muons, a scenario which is already tightly constrained by data from $B$-factories.

The dark photon couples to the right handed SM fermions at tree level with a coupling strength of $g_{T3R} = m_{A^'}/\sqrt{2}V$. The dark photon also has vector couplings to the other SM fermions through kinetic mixing  with coupling strength $\epsilon e$, where $\epsilon = g_{T3R} \sqrt{\alpha_{em}/ 4\pi^3}$. We assume  $\epsilon$ as a free parameter in our study. We consider $m_{A'} \le 2 m_{\mu}$ in order to avoid the bounds from BaBar~\cite{Aubert:2009cp, Lees:2014xha}. Therefore the possible final states of $A'$ decays are $\eta_{1,2} \eta_{2,1}$,$\nu \nu$, and $e^+ e^-$. Note that, only the last one is visible final state. There is no visible final state if $m_{A'} < 2m_e$, as $A' \rightarrow \gamma \gamma$ is forbidden by the Landau-Yang theorem~\cite{Landau:1948kw, Yang:1950rg}. In that scenario, either the $\eta_1 \eta_2$ or $\nu_S \nu_S$ will dominate the branching fraction, if allowed kinematically. If not, then the possible final states are $\nu_S \nu_A$, $\nu_A \nu_A$ but they are suppressed by the neutrino mixing angle.

The expression of the decay widths are\bea \Gamma_{\eta_1 \eta_2}^{A'} &=& \frac{m_{A^\prime}^3}{96 \pi V^2} \left(1 - \frac{4m_\eta^2}{m_{A'}^2} \right)^{1/2} \left(1 + \frac{2m_\eta^2}{m_{A'}^2} \right) ,\nonumber\\ \Gamma_{\nu_S \nu_S}^{A'} &=& \frac{m_{A'}^3}{ 12 \pi V^2} \left(1 - \frac{4m_{\nu_S}^2}{m_{A'}^2} \right)^{3/2} ,\nonumber\\\Gamma_{e^+ e^-}^{A'} &=& \frac{\epsilon^2 \alpha_{em} m_{A'}}{ 3}\left(1 - \frac{4m_e^2}{m_{A'}^2} \right)^{1/2} \left(1 + \frac{2m_e^2}{m_{A'}^2} \right). \eea

The possible visible final states of $\phi^'$ decay are $\mu^+ \mu^-$ and $\gamma \gamma$, via one loop. And the possible invisible final states are $\eta \eta$, $\nu \nu$, $A' A'$. If the produced $\nu_S$ or $A'$ decay to SM particles, they can also produce visible energy. If $m_{\phi'} > 2m_{A'}$, then $\phi'$ can decay promptly to $A'$. And if $m_{\phi'} > 2m_\pi$, then hadronic final states are possible at tree levels.  But the  branching fraction would be negligible compared to $\mu^+ \mu^-$, since the coupling to first-generation quarks is so small. The expressions for the decay widths are given by, \bea \Gamma^{\phi^\prime}_{A^\prime A^\prime} &=& \frac{m_{\phi^\prime}^3}{128\pi V^2}\left( 1-\frac{4m_{A^\prime}^2}{m_{\phi^\prime}^2}\right)^{1/2} \left(1+ 12\frac{m_{A^\prime}^4}{m_{\phi^\prime}^4}-4\frac{m_{A^\prime}^2}{m_{\phi^\prime}^2} \right),\nonumber\\ \Gamma_{\mu^+ \mu^-}^{\phi'} &=& \frac{m_\mu^2 m_{\phi'} }{16\pi V^2} \left(1 - \frac{4m_\mu^2}{m_{\phi'}^2} \right)^{3/2},\nonumber\\\Gamma_{\eta_i \eta_i}^{\phi'} &=& \frac{m_{\eta_i}^2 m_{\phi'} }{32\pi V^2} \left(1 - \frac{4m_{\eta_i}^2}{m_{\phi'}^2} \right)^{3/2},\nonumber\\\Gamma_{\nu_S \nu_A}^{\phi'} &=&  \frac{m_{\nu_D}^2 m_{\phi'} }{16\pi V^2} \left(1 - \frac{m_{\nu_S}^2}{m_{\phi'}^2} \right)^{2},\nonumber\\\Gamma_{\gamma \gamma}^{\phi'} &=& \frac{\alpha_{em}^2 m_\mu^4}{8\pi^3 m_{\phi'} V^2} \left[1 +  \left(1 - \frac{4m_\mu^2}{m_{\phi'}^2} \right) \left(\sin^{-1} \frac{m_{\phi'}}{2m_\mu} \right)^2 \right]^2,\nonumber\\\eea where we assume that $m_{\phi'} < 2m_\mu$ in order to calculate $\Gamma^{\phi^\prime}_{\gamma \gamma}$, otherwise, this decay would be negligible compared to the $\mu^+ \mu^-$ channel. Note that, $\phi'$ will always decay very promptly.

If the sterile neutrino mass, $m_{\nu_s} > 2 m_{\mu}$, then it decays promptly via $\nu_S \rightarrow \mu^+ \mu^- \nu_A$ at tree level. But if $m_{\nu_s} < 2 m_{\mu}$, then the following decay happens, $\nu_S \rightarrow \nu_A \gamma \gamma$. The rate is, \bea \Gamma_{\nu_S} &\propto& \alpha_{em}^2\frac{m_{\nu_S}^7 m_{\nu_D}^2}{m_{\phi'}^4 V^4} .\eea The lifetime can be estimated as, $\tau_{\nu_S} \sim {\mathcal{ O}}(10^{13})$~sec for $V=10$~GeV,  $m_{\phi'}\sim 100$~MeV,  $m_{\nu_S} = 10$~MeV,  $m_{\nu_D} = 10^{-3}$~MeV. Therefore, for the laboratory based experiments, they can be treated as stable particles. The decay $\nu_S \rightarrow \nu_A \gamma$ is also possible through a transition dipole at two-loop level but is highly suppressed.

The $U(1)_{T3R}$ has chiral couplings to the fermions. Due to this, the tree level production cross section of the longitudinal mode of $A'$ can be enhanced.  This makes $U(1)_{T3R}$ different compared to other well studied $U(1)$ gauge groups such as such as $U(1)_{B-L}$, $U(1)_{L_i-L_j}$, $U(1)_{X}$~\cite{Foot:1990mn, He:1990pn, He:1991qd, Borah:2020swo, Costa:2020krs}. For $U(1)_{T3R}$, we can get qualitatively new constraints. The enhancement in the production cross section entirely comes from the axial part of the chiral couplings. The pure vector part of the interactions vanishes identically for the longitudinal mode, using Ward identity. Therefore the enhancement in the production cross section only works if the $A'$ is produced at tree-level.  If $A'$ is produced through kinetic mixing, then the contribution from longitudinal polarization will again vanish identically due to the Ward Identity since this would be a pure vector interaction.

\subsection{A UV-completion } \label{sec:UV}

Although the fermions masses arise from a renormalizable operator in the effective field theory defined below the EWSB scale, in the field theory defined above this scale this same term must arise from the non-renormalizable operator  $1/\Lambda_fH \phi \bar f P_R f $, where $\lambda_f = \langle H \rangle / \Lambda_f$.   This operator can be arise from renormalizable operators in a UV-completion utilizing the universal seesaw mechanism~\cite{Berezhiani:1983hm,Chang:1986bp,Davidson:1987mh,DePace:1987iu,Rajpoot:1987fca,Babu:1988mw,Babu:1989rb,Babu:2018vrl} if we add a new set of vector-like heavy fermions $Q_f$, which are neutral under $U(1)_{T3R}$, and have the SM gauge charges of a right-handed fermion. We summarize the particles along with their charges in Table.~\ref{table:UVcharge}.

 \begin{table}[h]

\captionsetup{justification   = RaggedRight,
             labelfont = bf}
\caption{ \label{table:UVcharge}   The 
charges of the fields 
under the gauge groups of the model are shown. For the fermionic fields, we have shown the charges of  the left-handed component of each Weyl spinor.}
            
\centering
\begin{tabular}{ ll }
\hline\hline
~~~~~~~~~~~~~~~Particle  &$SU(3)_C\times SU(2)_L \times U(1)_Y \times U(1)_{T3R}$   \\\hline
~~~~~~~~~~~~~~~$\chi_{uL}$ & $(3,1, 2/3,0)$\\
~~~~~~~~~~~~~~~$\chi_{d L}$ & $(3,1, -1/3,0)$\\
~~~~~~~~~~~~~~~$\chi_{\mu L}$ & $(1,1,-1,0)$\\
~~~~~~~~~~~~~~~$\chi_{\nu L}$ & $(1,1,0,0)$\\
~~~~~~~~~~~~~~~$\chi_{uR}^c$ & $(3,1, -2/3,0)$\\
~~~~~~~~~~~~~~~$\chi_{d R}^c$ & $(3,1, 1/3,0)$\\
~~~~~~~~~~~~~~~$\chi_{\mu R}^c$ & $(1,1,1,0)$\\
~~~~~~~~~~~~~~~$\chi_{\nu R}^c$ & $(1,1,0,0)$\\
~~~~~~~~~~~~~~~$q_L$ & $(3,2,1/6,0)$\\
~~~~~~~~~~~~~~~$u_R^c$ & $(3,1,-2/3,-2)$\\
~~~~~~~~~~~~~~~$d_R^c$ & $(3,1,1/3,2)$\\
~~~~~~~~~~~~~~~$l_L$ & $(1,2,-1/2,0)$\\
~~~~~~~~~~~~~~~$\mu_R^c$ & $(1,1,1,2)$\\
~~~~~~~~~~~~~~~$\nu_R^c$ & $(1,1,0,-2)$\\
~~~~~~~~~~~~~~~$\eta_L$ & $(1,1,0,1)$\\
~~~~~~~~~~~~~~~$\eta_R^c$ & $(1,1,0,-1)$\\
~~~~~~~~~~~~~~~$H$ & $(1,2,1/2,0)$\\
~~~~~~~~~~~~~~~$\phi$ & $(1,1,0,2)$\\
\hline\hline
\end{tabular}
\end{table}

 At high scale, the renormalizable Lagrangian of the UV complete model can be written as, \bea -\mathcal{L}_{\text{Y}} &=& \lambda_{Lu} \bar{q}_L^\prime \chi_{uR}^\prime \tilde{H} + \lambda_{Ld} \bar{q}_L^\prime \chi_{dR}^\prime H   + \lambda_{L\nu} \bar{l}_L^\prime \chi_{\nu R}^\prime \tilde{H} + \lambda_{Ll} \bar{l}_L^\prime \chi_{\mu R}^\prime H  +\lambda_{Ru} \bar{\chi}_{uL}^\prime u_{R}^\prime \phi^*  +\lambda_{Rd} \bar{\chi}_{dL}^\prime d_{R}^\prime \phi \nonumber\\ &~& +\lambda_{R \nu} \bar{\chi}_{\nu L}^\prime \nu_{R}^\prime \phi^* +\lambda_{Rl} \bar{\chi}_{\mu L}^\prime \mu_{R}^\prime \phi + m_{\chi_u} \bar{\chi}_{uL}^\prime \chi_{uR} + m_{\chi_d} \bar{\chi}_{dL}^\prime \chi_{dR} +m_{\chi_\nu} \bar{\chi}_{\nu L}^\prime \chi_{\nu R} + m_{\chi_\mu} \bar{\chi}_{\mu L}^\prime \chi_{\mu R} \nonumber \\ &~&+ m_D \bar{\eta}_R \eta_L + \frac{1}{2} \lambda_{\eta L} \bar{\eta}^c_L \eta_L \phi  + \frac{1}{2} \lambda_{\eta R} \bar{\eta}^c_R \eta_R \phi^* + H.c.~,~\, \eea

 In the flavor basis, the fermionic mass matrix can be written as,  \bea M_f = \left( \begin{array}{cc} 0 & \frac{\lambda_{Lf} v}{\sqrt{2}} \\ \lambda_{Rf} V & m_{\chi^\prime_f} \end{array} \right). \eea This can be diagonalized using seesaw mechanism leading to two mass eigenstates. The lightest of them is the SM fermion with mass \be m_f  = \frac{\lambda_{Lf} \lambda_{Rf} vV}{ \sqrt{2} m_{\chi^\prime_f}}, \ee  while the heavier one is the physical vector-like fermion with mass \be m_{\chi_f} \simeq m_{\chi^\prime_f}.  \ee They can be expressed in terms of the flavor eigenstates as, \begin{equation} \left( \begin{array}{c} f_{L,R} \\ \chi_{f_{L,R}} \end{array} \right) =  \left( \begin{array}{cc} \cos \theta_{f_{L,R}} & \sin  \theta_{f_{L,R}} \\ -\sin  \theta_{f_{L,R}} &\cos  \theta_{f_{L,R}} \end{array} \right) \left( \begin{array}{c} f^\prime_{L,R} \\ \chi_{f^\prime_{L,R}} \end{array} \right) ~,~\,  \end{equation}  where $\theta_{f_{L,R}}$ are the mixing angles.
 
We may then write  \bea{\cal L} &=& -\tilde m_{\chi_f} \bar \chi_f \chi_f - \lambda_{Lf} H^{(*)} \bar \chi_{fR} f_L - \lambda_{Rf} \phi^{(*)} \bar \chi_{fL} f_R +h.~c.,\eea where $f_R$ is a right-handed fermion charged under $U(1)_{T3R}$, and $f_L$ is the corresponding $SU(2)_L$ doublet containing the left-handed fermion.  Note that,  for $\lambda_{Lf,Rf} \sim {\cal O}(1)$, we would need $\tilde m_{\chi_\mu} \sim {\cal O}(10\tev)$, which is beyond the range of the LHC, but potentially within reach of the next generation of energy-frontier colliders. 

\subsection{Distinction between $U(1)_{T3R}$ and $U(1)_{B-L}$} \label{sec:BLT3R}

It is sometimes thought that the $U(1)_{T3R}$ scenario is 
a subspecies of $U(1)_{B-L}$, because one can express the 
hypercharge of SM fermions as $Y = Q_{T3R} + (1/2) Q_{B-L}$, 
for an appropriate normalization of the SM fermion $U(1)_{T3R}$ 
charges.  In this case, gauging $U(1)_Y$ and $U(1)_{B-L}$ is 
equivalent to gauging $U(1)_{T3R}$.  

But more generally, one can express the hypercharge as 
$Y = Q_{T3R} + (1/2) Q_{B-L} +Q_G$, where $Q_G$ is the charge 
under gauge group $U(1)_G$, under which SM fermions are 
neutral ($Q_G=0$).  But although SM fermions are uncharged under 
$U(1)_G$, the charge of the SM Higgs is an open question.  
Since the Standard Model Higgs doublet has hypercharge 
$Y=1/2$, but is neutral under $U(1)_{B-L}$, it must be 
charged under $U(1)_{T3R}$ and/or $U(1)_G$.  In the original 
$U(1)_{T3R}$ scenario, the SM Higgs was taken to be charged 
under $U(1)_{T3R}$, but a singlet under $U(1)_G$.  In this case, 
$U(1)_G$ decouples from the SM at tree-level, and this scenario 
indeed is related to the gauging of $U(1)_{B-L}$, along with 
(potentially) a secluded $U(1)$.  But in this case, condensation 
of the SM Higgs breaks both $U(1)_Y$ and $U(1)_{T3R}$, and one 
cannot decouple the two symmetry-breaking scales.

In the more recently studied version of this scenario, the 
SM Higgs is neutral under $U(1)_{T3R}$.  Thus, if one wishes to 
connect hypercharge to $U(1)_{B-L}$ and $U(1)_{T3R}$, one must 
couple the SM Higgs to  $U(1)_G$.  The symmetry-breaking 
scale of $U(1)_{T3R}$ is now decoupled from the electroweak 
symmetry-breaking scale, and is controlled by the dark Higgs, 
which is charged under $U(1)_{T3R}$ and $U(1)_G$, but is a singlet 
under hypercharge.  Note that it is not necessary for there to exist 
an additional gauge group $U(1)_G$, but that if one is not 
present, then one cannot express hypercharge in terms of 
$U(1)_{B-L}$ and $U(1)_{T3R}$, given the charges of the SM 
Higgs and dark Higgs.

Thus, the distinction between the most recent incarnations of 
$U(1)_{T3R}$ and previous scenarios (including $U(1)_{B-L}$) lies 
in the decoupling of the $U(1)_{T3R}$ and electroweak symmetry-breaking scales, and resulting introduction of a new 
scalar mediator.

\section{Constraints on the paramter space} \label{sec:constraints}

There is a large literature discussing laboratory, astrophysical, and cosmological constraints on models with a light dark photon or dark Higgs  (for example, see~\cite{Batell:2017kty,Bauer:2018onh}).  But if the new gauge group is $U(1)_{T3R}$, then there are some qualitatively different constraints~\cite{Dutta:2020jsy}.  We discuss various constraints that are applicable for the scenario.

\begin{itemize}

\begin{figure}[h]
\centering
\includegraphics[height=10cm,width=14.2cm]{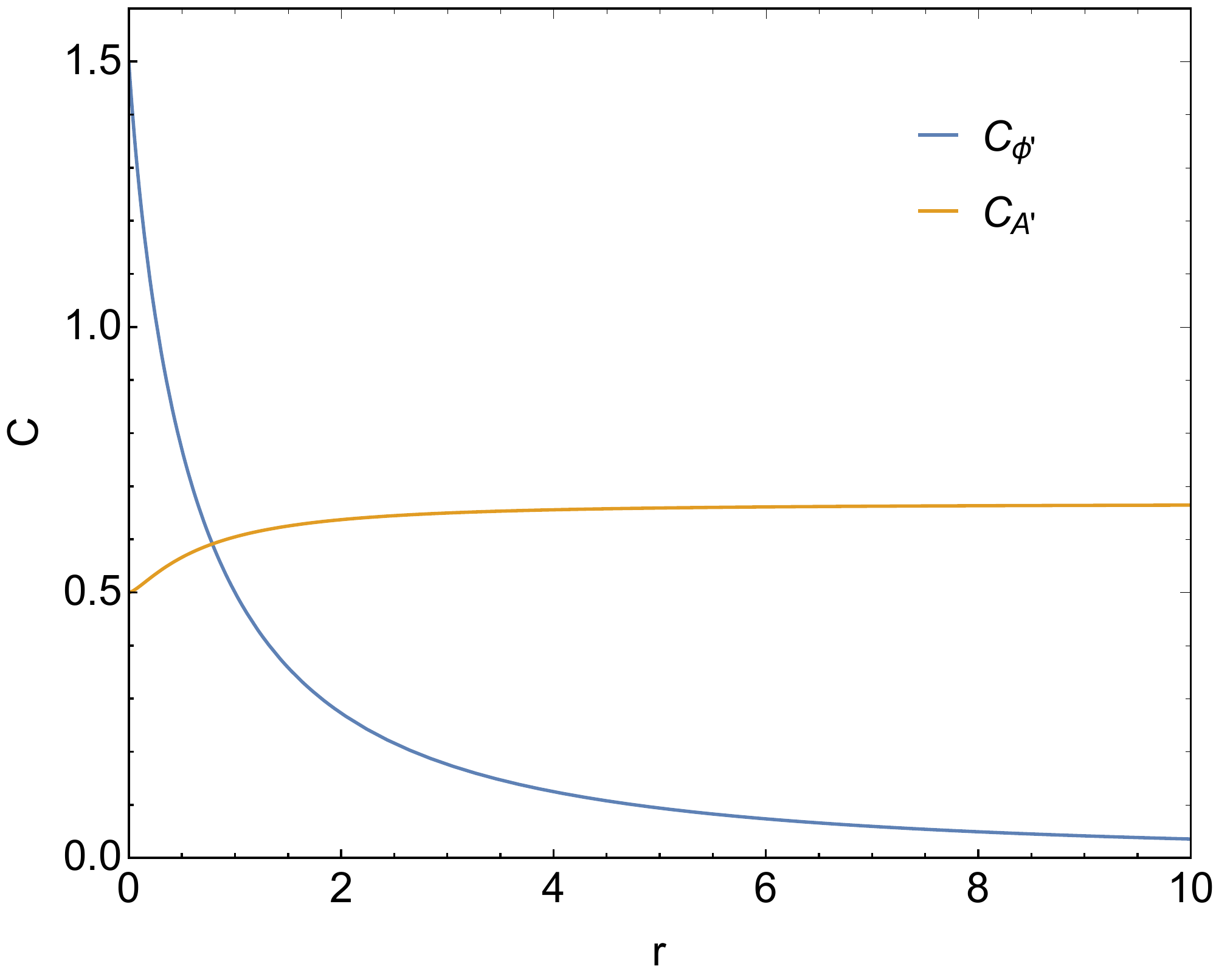}
\captionsetup{justification   = RaggedRight,
             labelfont = bf}
\caption{\label{fig:Cvsrplot}  
Plot from Ref.~\cite{Dutta:2021afo}: variation of $C_{\phi'}$ and $C_{A'}$ as functions of $r_{\phi'}$ and 
$r_{A'}$, respectively.}
\end{figure}

\item {\it Cosmological and Astrophysical Observables:} Because the dark photon couples to right-handed SM fermions, the longitudinal polarization does not decouple from tree-level processes.  This yields an enhanced cross section for any process in which a hard dark photon is produced from a tree-level process.  As an example, we can consider the scenario where the muon is the only charged lepton coupling to the dark photon. This scenario is typically subject to much weaker constraints.   But it has  been shown that, if the Universe reheats to a sufficiently high temperature ($\gtrsim 0.1\gev$), the coupling of the dark photon to right-handed muons would lead to enhanced production of the dark photon in the early Universe;constraints on $\Delta N_{eff}$ thus rule out such scenarios for $m_{A'} \lesssim 1\mev$ for {\it arbitrarily small coupling} unless the symmetry-breaking scale is $> {\cal O}(10^6)\gev$~\cite{Dutta:2020jsy}.  Recent astrophysical constraints on ALPs coupling to muons in supernovae~\cite{Bollig:2020xdr,Croon:2020lrf} can easily be repurposed as constraints on the longitudinal polarization of the dark photon (equivalently, the Goldstone mode), and these constraints are comparable.  This constraints together place tight bounds on scenarios with $m_{A'} \lesssim 1\mev$.

\item {\it Anomalous magnetic moment of muon}: The correction to $a_\mu \equiv (g_\mu -2)/2$ in this model is given  by~\cite{Dutta:2021afo} \bea \delta a_\mu &=& (6.98 \times 10^{-7}) \left( \frac{V}{10\gev} \right)^{-2} \left(C_{\phi'} - C_{A'} \right) , \eea where  \bea C_{\phi'} &=& \int_0^1 dx \frac{(1-x)^2(1+x)}{(1-x)^2+x r_{\phi'}^2} ,\nonumber\\ C_{A'} &=&   \int_0^1 dx \frac{x(1-x)(2-x) r_{A'}^2 +x^3}{x^2+(1-x)r_{A^\prime}^2} , \eea are the contributions from one-loop diagrams with  the $\phi'$ and $A'$ in the loop, respectively, and $r_{\phi'} \equiv m_{\phi'}/m_\mu$,  $r_{A'} \equiv m_{A'}/m_\mu$.  Note that the contributions of $\phi'$ and $A'$ are necessarily of opposite sign, because the $A'$ has both vector and axial couplings to the muon. The variation of $C_{\phi^', A^'}$ versus $r$ is shown in Fig.~\ref{fig:Cvsrplot}.
\begin{figure}[h]
\centering
\includegraphics[height=10cm,width=14.2cm]{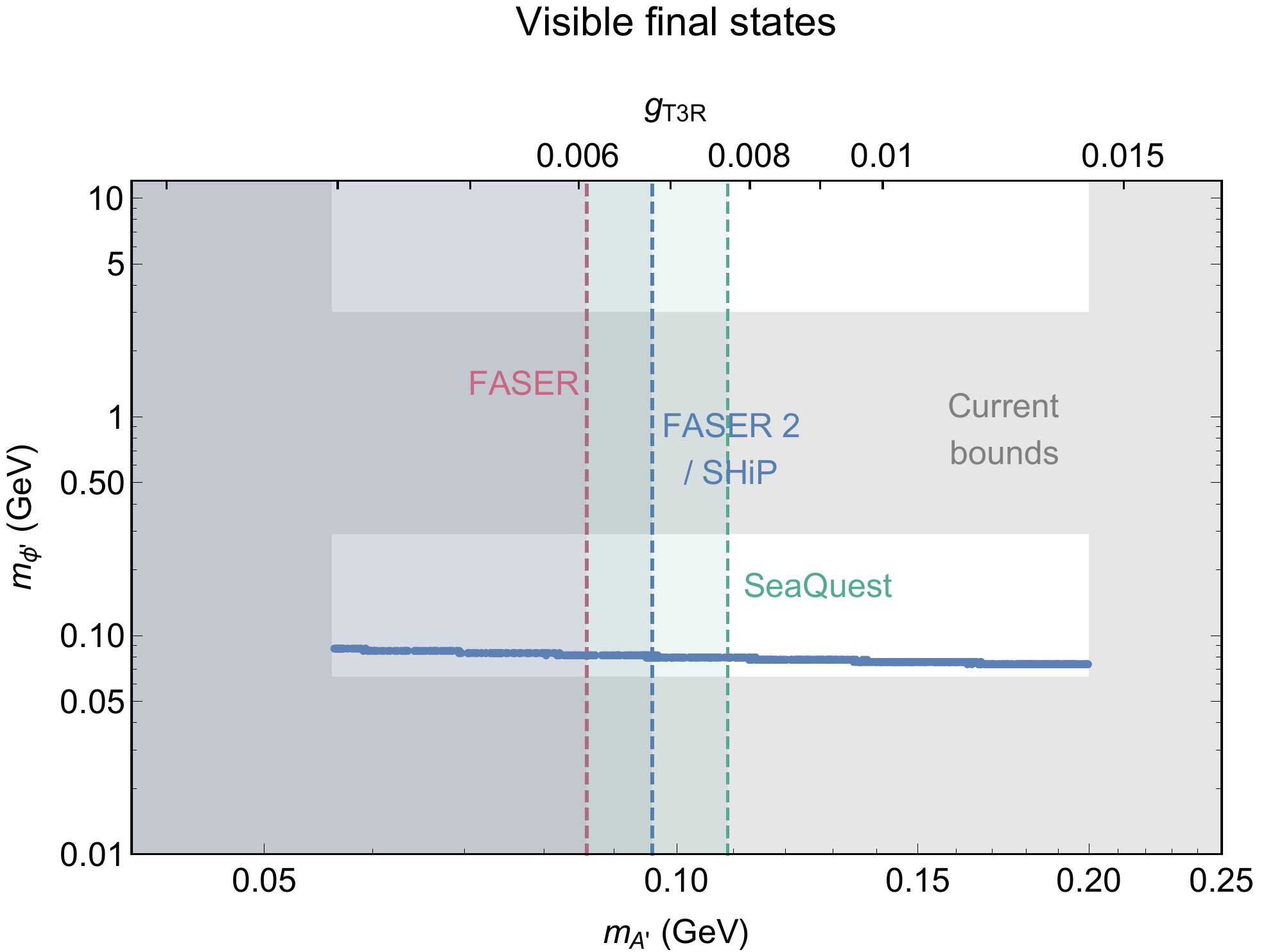}
\captionsetup{justification   = RaggedRight,
             labelfont = bf}
\caption{\label{fig:visiblefuture}  
Plot from Ref.~\cite{Dutta:2021afo} of the region in 
the $(m_{A'}, m_{\phi'})$-plane which is consistent with current 
measurements of $g_\mu-2$ (blue), along with current exclusion bounds 
(grey) from U70/NuCal~\cite{Gninenko:2014pea, Davier:1989wz, Bauer:2018onh}, E137~\cite{Riordan:1987aw,Bjorken:1988as, Bjorken:2009mm}, Orsay~\cite{Davier:1989wz, Bauer:2018onh}, and Babar~\cite{Aubert:2009cp, Lees:2014xha, Bauer:2018onh}, and the future sensitivity of 
FASER~\cite{Feng:2017uoz, Ariga:2018zuc, Ariga:2018pin, Ariga:2018uku, Ariga:2019ufm} (red transparent), FASER 2/SHiP~\cite{Anelli:2015pba, Alekhin:2015byh} (blue transparent) and SeaQuest~\cite{Berlin:2018pwi, Aidala:2017ofy} (green 
transparent).  $g_{T3R}$ is shown on the top axis.}
\end{figure}

Interestingly, the contribution of the $A'$ diagram to $\delta a_\mu$ is nearly universal, with $C_{A'}$ confined to lie between $1/2$ and $2/3$.  In particular, even at small coupling ($m_{A'} / V \ll 1$), although the transverse polarizations decouple, the contribution of the longitudinal mode remains unsuppressed; it becomes essentially the Goldstone mode, as expected from the Goldstone Equivalence theorem.  Moreover, if $m_{\phi'} \lesssim m_\mu$, then $C_{\phi'}$ is also an ${\cal O}(1)$ number.  This also is a result of the Goldstone Equivalence theorem.  $C_{A'}$ varies only slightly, but for small $m_{A'}$, $C_{A'}$ receives contributions only from the Goldstone mode $\sigma$.  Since the $\sigma$ and $\phi'$ have the same coupling, $C_{A'}$ and $C_{\phi'}$ must be comparable in magnitude and opposite in sign.  If they cancel to within ${\cal O}(1\%)$, then this model is consistent with measurements of $g_\mu -2$.  Interestingly, this cancellation occurs in a region of parameter space which is not excluded by current experiments, but which can be probed by experiments at the Forward Physics Facility (FPF)~\cite{Feng:2022inv}.

\begin{figure}[th]
\centering
\includegraphics[height=10.5cm,width=14.2cm]{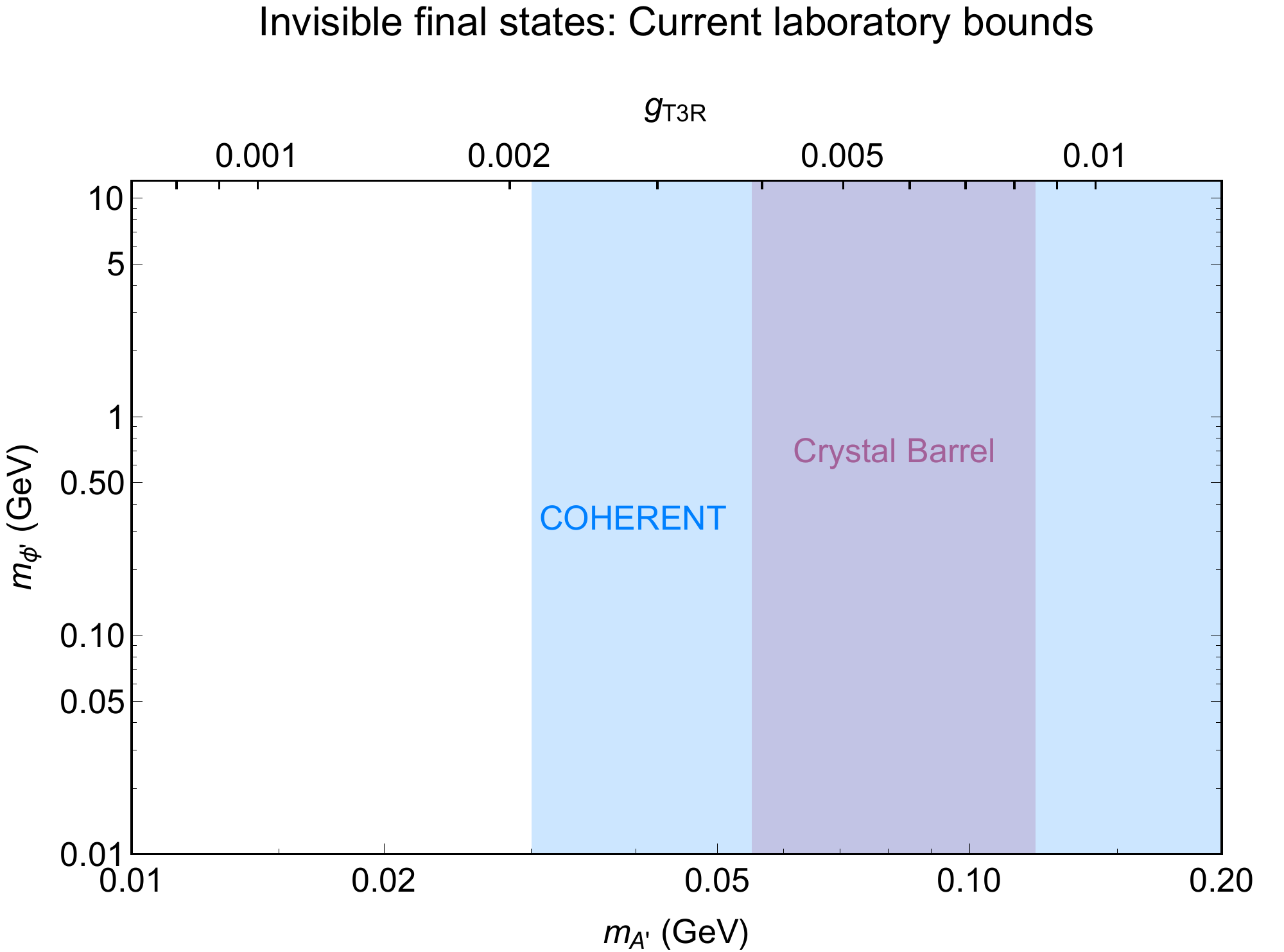}

\caption{\label{fig:invisible} Plot from Ref.~\cite{Dutta:2020jsy} of the region in 
the $(m_{A'}, m_{\phi'})$-plane which is  excluded by current 
laboratory experiments, assuming that $A'$ and 
$\phi'$ decay dominantly to invisible final states.  
Included are bounds from COHERENT~\cite{Akimov:2017ade, Akimov:2018vzs, Akimov:2018ghi, Akimov:2019xdj, Akimov:2020pdx}(light blue) and Crystal Barrels~\cite{Amsler:1994gt, Amsler:1996hb} (light purple) experiments.  $g_{T3R}$ is shown on the top axis.}
\end{figure}

\item {\it Visible Decay}: We will focus here on the case in which the mediators $A'$ and $\phi'$ decay dominantly through the visible channels $A' \rightarrow e^+ e^-$ (the $\gamma \gamma$ channel is forbidden by the Landau-Yang theorem) or $\phi' \rightarrow \gamma \gamma$.  The decay widths for these processes are \bea \Gamma_{A' \rightarrow e^+ e^-} &=& \frac{\epsilon \alpha_{em} m_{A'} }{3} \left(1-\frac{4m_e^2}{m_{A'}^2} \right)^{-1/2} \left(1+\frac{2m_e^2}{m_{A'}^2} \right), \nonumber\\\Gamma_{\phi' \rightarrow \gamma \gamma} &=& \frac{\alpha_{em}^2 m_\mu^4}{4\pi^3 V^2 m_{\phi'}} \left[1+\left(1-\frac{4m_\mu^2}{m_{\phi'}^2} \right)\left(\sin^{-1} \frac{m_{\phi'}}{2m_\mu} \right)^2\right]^2,\eea where $\epsilon$ parameterizes the kinetic mixing between $U(1)_{T3R}$ and $U(1)_{em}$. In general, $\epsilon$ is a free parameter.We will take $\epsilon = g_{T3R} \sqrt{\alpha_{em} / 4\pi^3}$, which is the magnitude of the contribution one would get from a one-loop diagram involving a right-handed fermion.

Note that fixing $m_{A'}$, $m_{\phi'}$ and $V$ is sufficient to specify the mediator couplings, production cross sections, and decay rates.  In Figure.~\ref{fig:visiblefuture}, we plot bounds on this scenario in the $(m_{A'}, m_{\phi'})$-plane, setting $V = 10~\gev$ as a benchmark.  We see that for $m_{A', \phi'} > 2m_\mu$, this scenario is tightly constrained by searches at Babar~\cite{Aubert:2009cp, Lees:2014xha, Bauer:2018onh} for prompt decays of the mediators to muons.

But we see that for $m_{A',\phi'} < 2m_\mu$, there is open parameter space.~\footnote{There is also open parameter space when $\phi'$ is heavy enough that production at $B$-factories is suppressed.}  This region of parameter space lies above the ``ceiling" of current displaced detector searches.The reason, essentially, is that the lifetime of the mediators decreases with increasing mediator mass.  Below the threshold for tree-level decay, the lifetime of the mediators is long enough for them to escape near detectors, but still short enough that they decay before reaching displaced detectors.  This window remains open until the decay lengths become long enough for the particles to reach existing displaced detectors, with leading current bounds being set by U70/NuCal~\cite{Gninenko:2014pea, Davier:1989wz, Bauer:2018onh} (in the case of $A'$) and E137~\cite{Riordan:1987aw,Bjorken:1988as, Bjorken:2009mm} (in the case of $\phi'$). Importantly, this open window includes the region in which constraints on $g_\mu -2$ are also satisfied.New instruments at FPF can probe this open window.

We focus here on sensitivity to the dark photon.An estimated sensitivity of experiments at FPF to this scenario can be extrapolated from a sensitivity to that of a secluded $U(1)$ as described in Ref.~\cite{Dutta:2020enk},  These sensitivities are plotted in Figure.~\ref{fig:visiblefuture}.

\item {\it Invisible Decay:}  If the mediators decay invisibly (either to neutrinos or dark matter), then there is a complementary set of detection possibilities.  If $m_{A'} \lesssim 10~\mev$this scenario is constrained by measurements of $N_{eff}$ at the time of recombination~\cite{Dutta:2020jsy}.  For $m_{A'} \gtrsim 10~\mev$, this scenario can be ruled out by measurements from COHERENT~\cite{COHERENT:2017ipa,COHERENT:2018imc,COHERENT:2018gft,COHERENT:2019kwz,COHERENT:2020iec} (which can search for the scattering of long-lived particles at a displaced detector) and Crystal Barrel~\cite{AMSLER1994271,CrystalBarrel:1996xfs} (which searches for anomalous $\pi^0$ decay).  But for $m_{A'} \sim 10~\mev$, this scenario could explain an anomalous event rate seen at COHERENT~\cite{Dutta:2019nbn}.  In this scenario, the COHERENT anomalous event rate arises from the decay of $A'$ into either dark matter or right-handed neutrinos, with the invisible particles subsequently scattering at the COHERENT detector.

But upcoming experiments which search for new particles which decay invisibly (via missing energy signatures) can probe all of the available parameter space for this scenario.  In particular, the focus is on experiments which create a muon beam, such as NA-64$\mu$~\cite{Chen:2017awl,Gninenko:2019qiv} and LDMX-M${}^3$~\cite{Kahn:2018cqs,Berlin:2018bsc}.  Because the longitudinal polarization of the gauge boson necessarily has a large coupling to the muon, the $A'$ will be produced copiously, and either of these experiments would be capable of excluding all of the available parameter space, in the case of invisible decays. The corresponding bounds are shown in Fig.~\ref{fig:invisible}.

\end{itemize}

The summary of all the constraints are presented  in Table.~\ref{table:result}.

\section{Direct detection} \label{sec:direct detection}

  Direct detection experiments can play an especially interesting role in proving the presence the dark matter in the universe. In the direct detection experiments, the dark matter is supposed to hit the target material and generate recoil in the nucleus, which is detected as deposited energy. This idea does not work best for sub-GeV dark matter as the traditional direct detection experiments lose sensitivity below a certain recoil energy. Recently a plethora of ideas have been introduced to perform direct detection experiment for sub-GeV dark matter. A few such experiments already give bounds on the dark matter-nucleon scattering cross section. We summarize them here. 
  
  \begin{itemize}
   \item XENON1T: If the slowly moving dark matter particles get boosted by the interactions with cosmic ray, they can deposit enough recoil energy to be detected by the detector. Current data gives a bound $\sigma_{SI} \le \mathcal{O} (10^{-29}-10^{-30})$~cm$^2$  or $\sigma_{SI} \ge \mathcal{O} (10^{-28})$~cm$^2$~\cite{Bringmann:2018cvk, Tooth:2019krz}.
   \item CRESST-III: Can put bounds on dark matter-nucleon scattering cross section for dark matter mass as low as 200 MeV. The cross section has to be less than $\sigma_{SI} \sim 10^{-35}$~cm$^2$ ~\cite{CRESST:2019jnq}.
   \item CDEX-1B: This experiment gives bounds for the mass range 50-180 MeV utilising the Migdal effect. The cross section is required to be less than $\sigma_{SI} \sim 10^{-32}- 10^{-34}$~cm$^2$~\cite{CDEX:2019hzn}.
  	
  \end{itemize}

We find that the parameter space of our model satisfies all these bounds. The projection for the differential event rates for future experiments with low thresholds are also very interesting. Due to the presence two new light mediators, we have two distinct mode for generating spin-independent(SI) dark matter-nucleon scattering process.

\begin{itemize}
	\item {\it Mediated by $\phi^'$}: This process is elastic, SI and isospin-invariant. The dark matter-nucleon scattering cross section at zero momentum trasnfer is given by, \begin{equation}							\sigma_{SI}^{scalar(p,n)} = \frac{\mu_{\eta N}^2 m_\eta^2}{4\pi V^4 m_{\phi'}^4} f_{p,n}^2 \end{equation} where ~\cite{Falk:1999mq},
\begin{equation} \frac{f_{p,n}}{m_N}=\sum_{q=u,d,s}f_{T_q}^{(p,n)}\frac{f_q}{m_q}+\frac{2}{27}\left(1-\sum_{q=u,d,s}f_{T_q}^{(p,n)}\right) \sum_{q=c,b,t}\frac{f_q}{m_q}. \end{equation} The values of different quantities are: $f_{u,d} = m_{u,d}$, $f_{s,c,b,t}=0$;  $f_{T_u}^{(p)}$, $f_{T_d}^{(p)}$ and $f_{T_s}^{(p)}$ are  0.019, 0.041 and 0.14, respectively~\cite{Gasser:1990ce}; and  $f_{T_u}^{(n)}$, $f_{T_d}^{(n)}$ and $f_{T_s}^{(n)}$ are  0.023, 0.034 and 0.14, respectively~\cite{Gasser:1990ce}. The threshold velocity as a function of the nuclear recoil can be written as, \begin{equation} \label{vminel} v_{min} = \frac{\sqrt{2m_A E_R}}{2 \mu_{\eta A}} . \end{equation}

	\item {\it Mediated by $A^'$}: Since the dark photon couples to up-type and down-type fermions with opposite charge, it leads to isospin-violating~\cite{Chang:2010yk, Feng:2011vu, Feng:2013vod} spin-independent scattering.  Moreover, since the dark matter candidate(s) are Majorana fermions, one necessarily has inelastic scattering.  Indeed, scattering via a dark photon is necessarily inelastic 
	if the dark matter is charged only under spontaneously-broken continuous symmetries; in that case, the dark matter is generically 
	a real degree of freedom, which cannot couple through a diagonal vector current. In this case, the dark matter-nucleon scattering cross section at zero momentum transfer is given by, 
	\bea  
	\sigma_{SI}^{vector(p, n)} &=& 
	\frac{\mu_{\eta N}^2}{16\pi V^4} .  
	\eea   
	The threshold velocity is given by, 
	\begin{equation} 
	v_{min} = \frac{1}{\sqrt{2m_A E_R}}\left( \frac{m_A E_R}{\mu_{\eta A}}+\delta \right) ,  
	\end{equation} 
	where we only keep terms linear in $\delta$, considering small $\delta$. In this limit we have, $\mu_{\eta_j N}\simeq\mu_{\eta_i N}=\mu_{\eta N}$.
\end{itemize}

\begin{figure}[h]
\begin{subfigure}[b]{0.485\textwidth}
\includegraphics[width=1.0\linewidth,height=6cm]{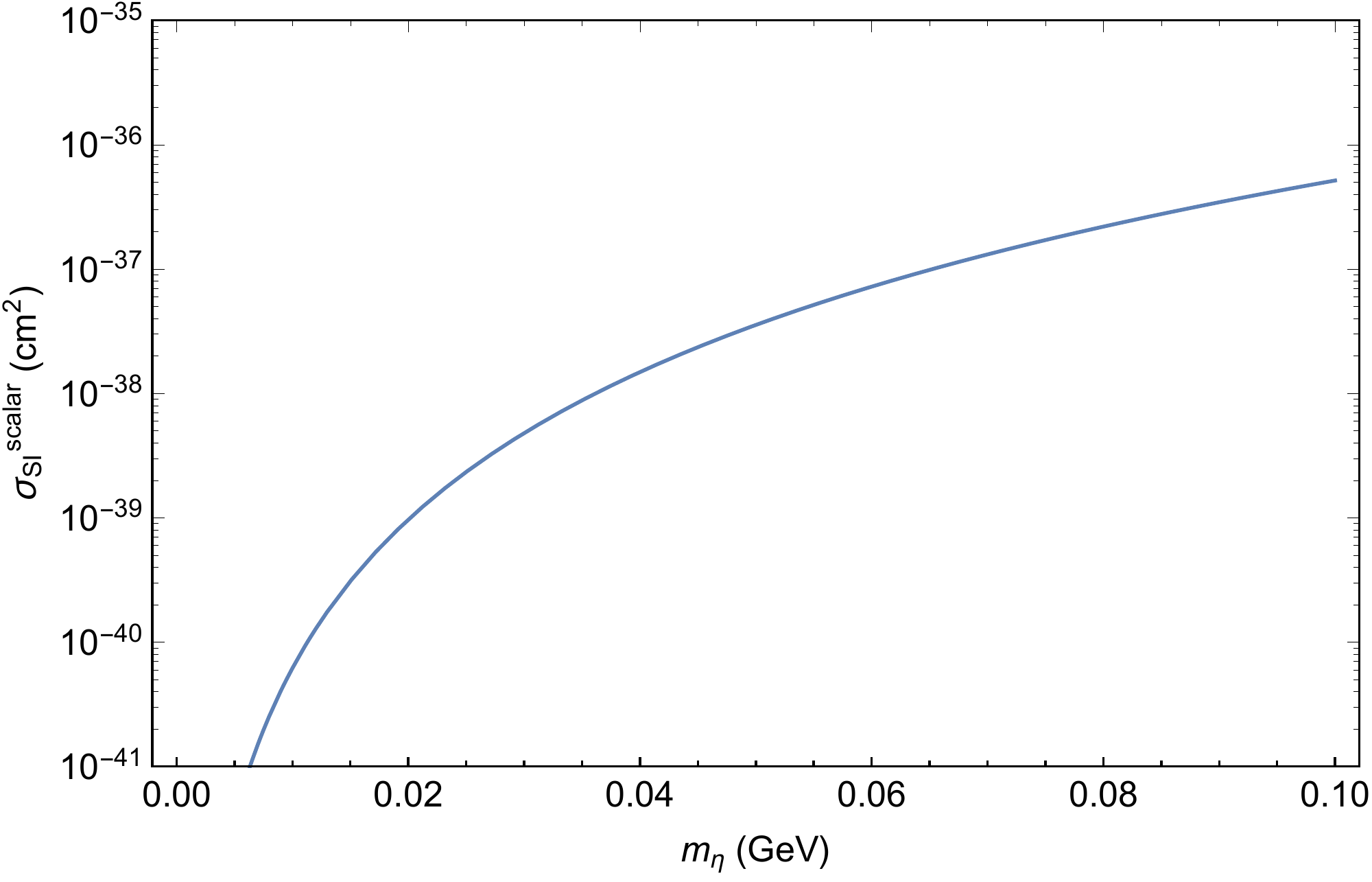}
\captionsetup{labelfont = bf}
\caption{\label{fig:scalarcrosssection}}
\end{subfigure}	
\hspace{0.2cm}	
\begin{subfigure}[b]{0.485\textwidth}
\includegraphics[width=1.0\linewidth,height=6cm]{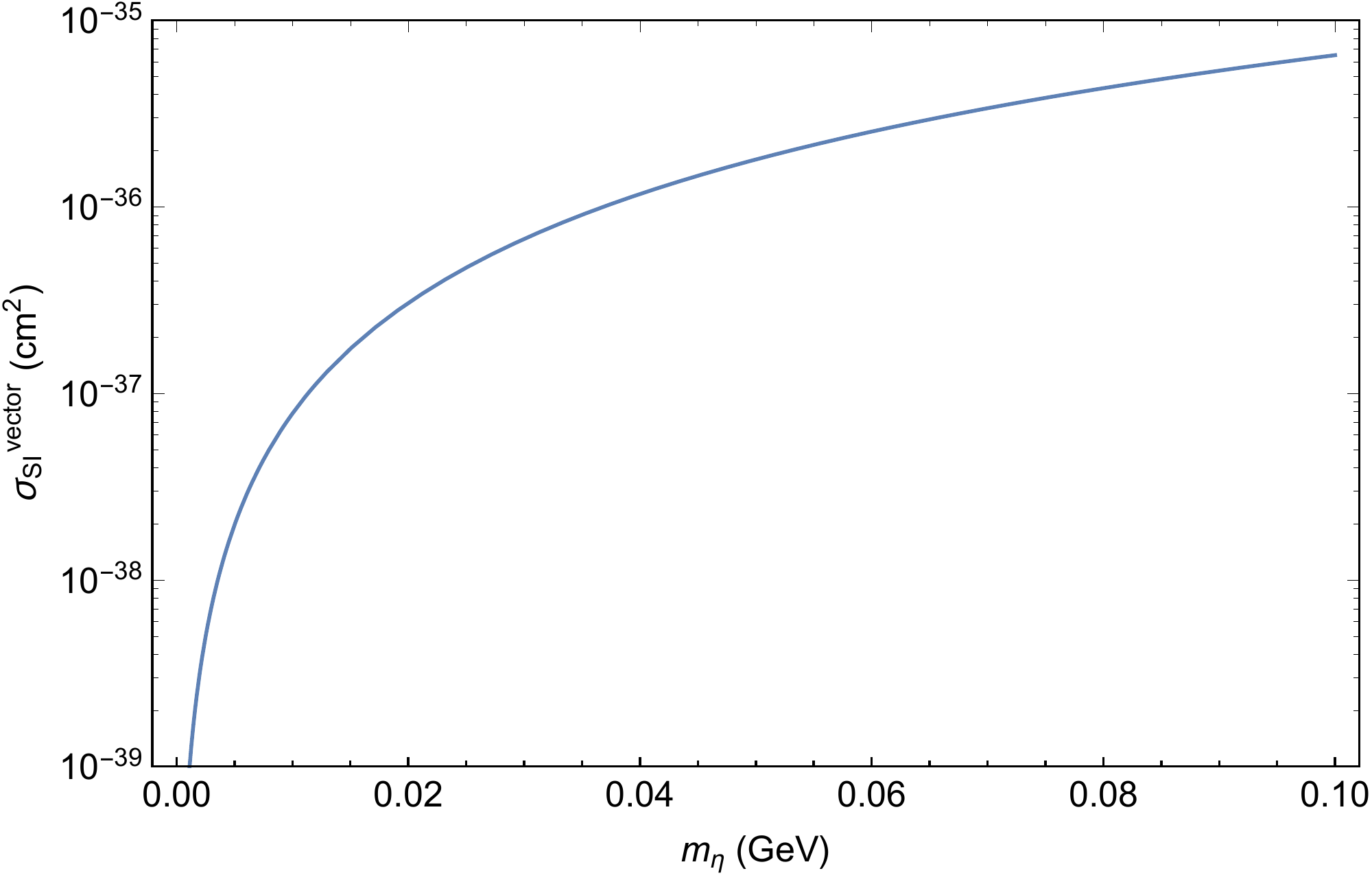}	
\captionsetup{labelfont = bf}
\caption{\label{fig:vectorcrosssection}}
\end{subfigure}	
\captionsetup{justification   = RaggedRight,
             labelfont = bf}
\caption{\label{fig:crosssection} Dark matter-nucleon scattering cross section is shown  as a function of the dark matter masses. We assume $m_{\phi^\prime}=$ 200 MeV, $\delta =0$ and $V=$ 10 GeV. } 
\end{figure}

\begin{figure}[h]
\begin{subfigure}[b]{0.485\textwidth}
\includegraphics[width=1.0\linewidth,height=6cm]{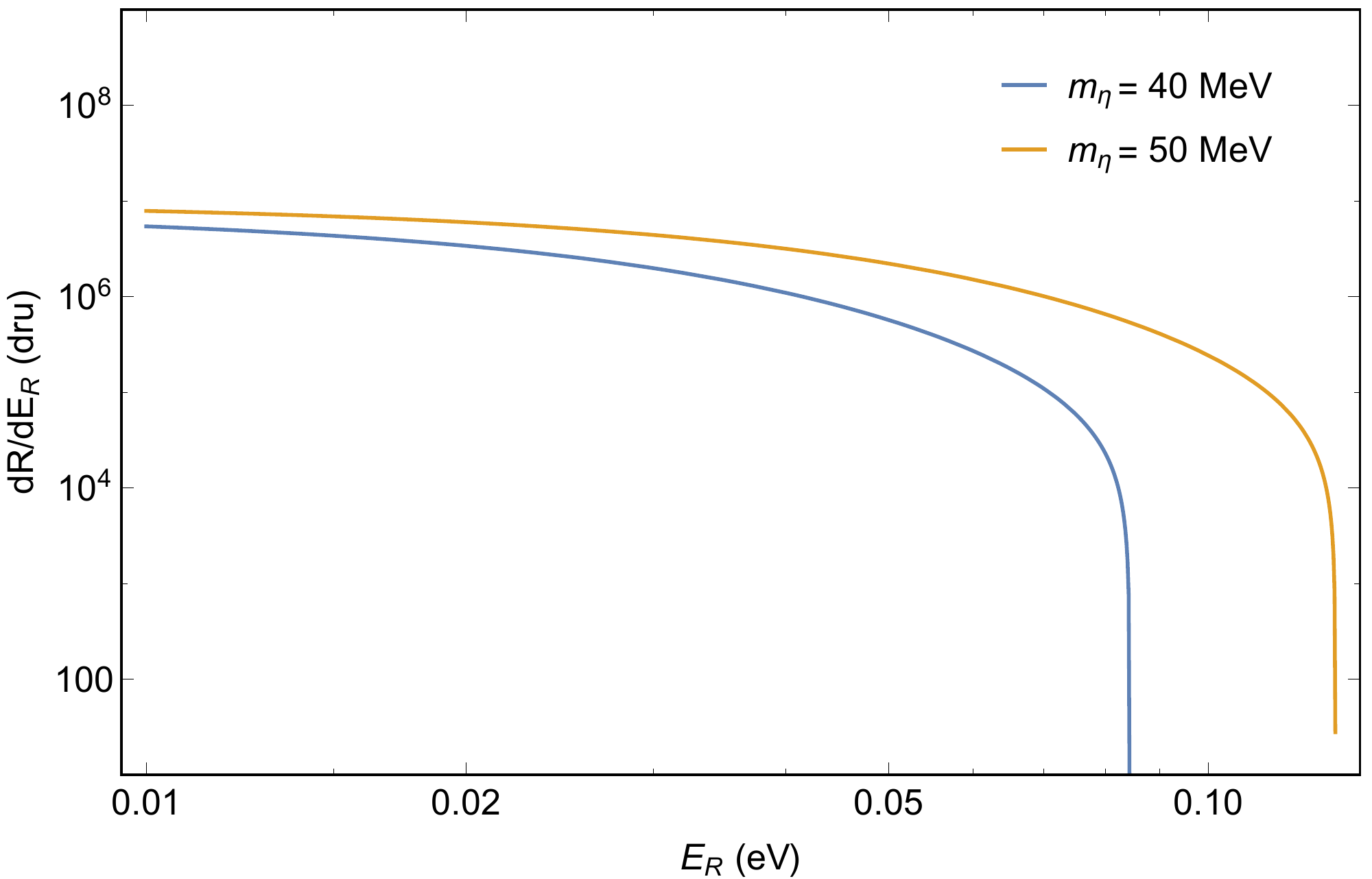}
\captionsetup{labelfont = bf}
\caption{\label{fig:elastic01}}
\end{subfigure}	
\hspace{0.2cm}	
\begin{subfigure}[b]{0.485\textwidth}
\includegraphics[width=1.0\linewidth,height=6cm]{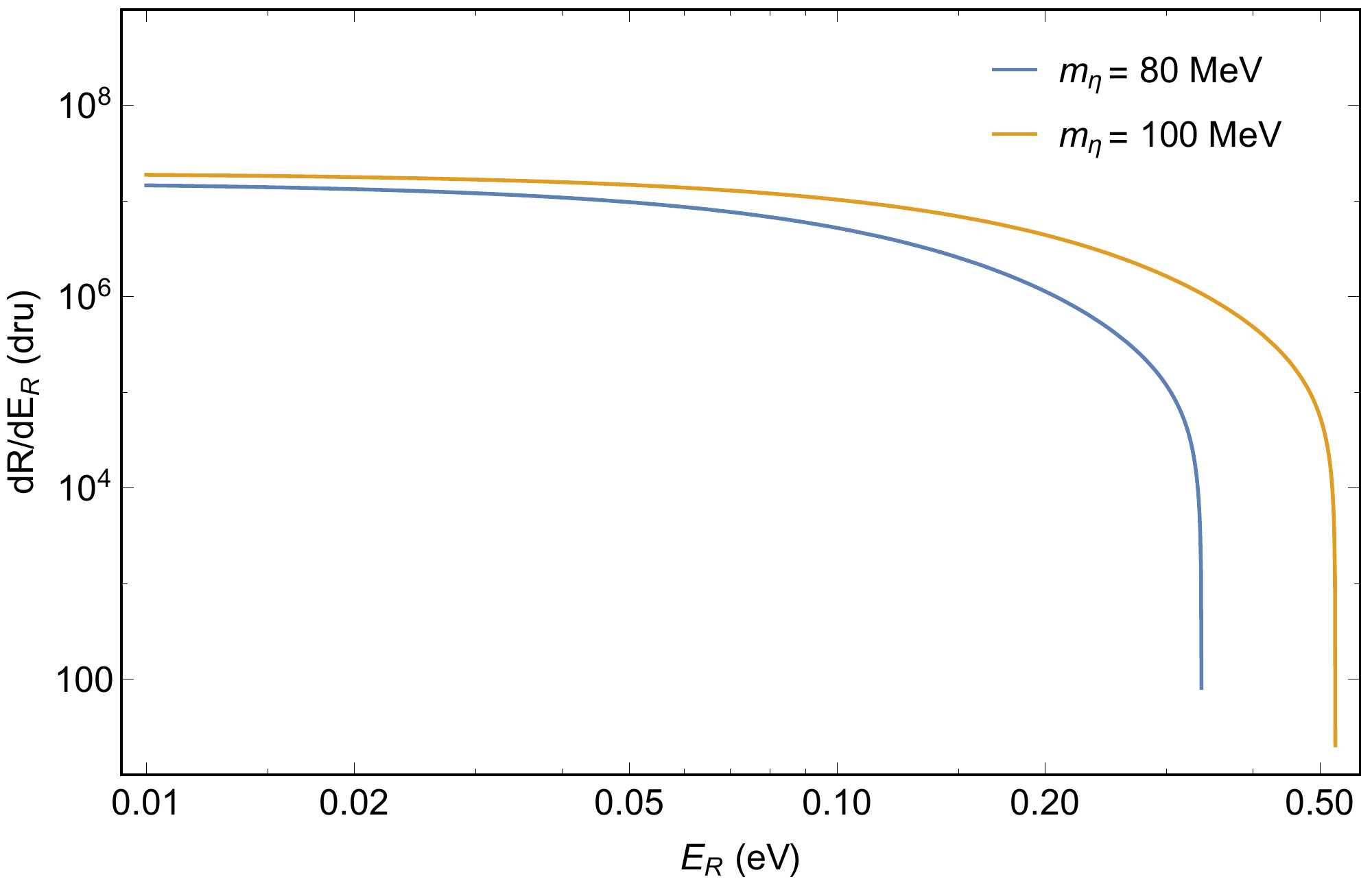}	
\captionsetup{labelfont = bf}
\caption{\label{fig:elastic02}}
\end{subfigure}	
\captionsetup{justification   = RaggedRight,
             labelfont = bf}
\caption{\label{fig:elastic}  Recoil spectrum   for elastic scattering off a Xenon nucleus for different dark matter masses. We used, $m_{\phi^\prime}=$ 200 MeV and $V=$ 10 GeV. Note that, the upper limit of recoil energy increases with increasing dark matter masses. } 
\end{figure}

\begin{figure}[h]
\begin{subfigure}[b]{0.485\textwidth}
\includegraphics[width=1.0\linewidth,height=6cm]{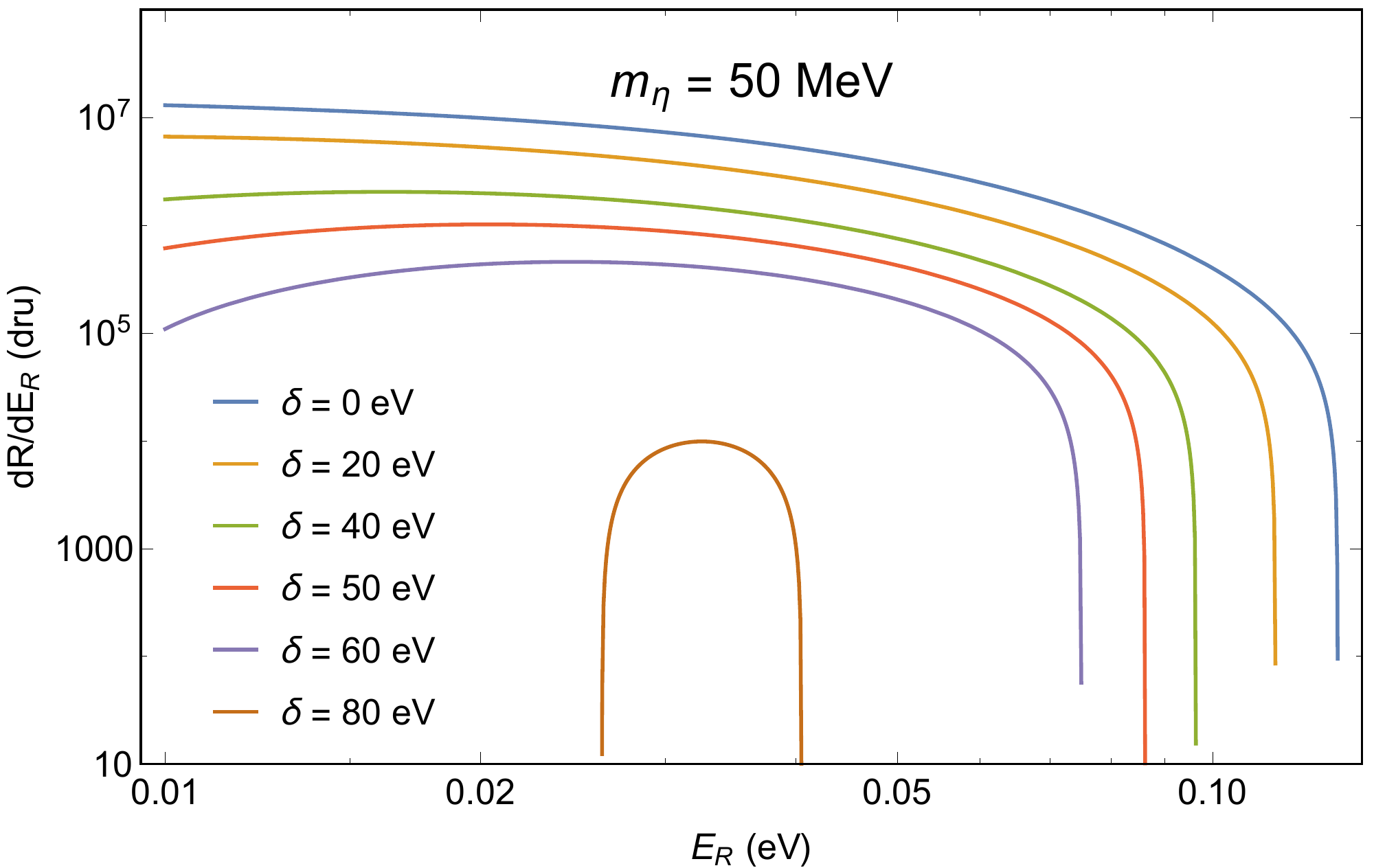}
\captionsetup{labelfont = bf}
\caption{\label{fig:inelastic01}}
\end{subfigure}	
\hspace{0.2cm}	
\begin{subfigure}[b]{0.485\textwidth}
\includegraphics[width=1.0\linewidth,height=6cm]{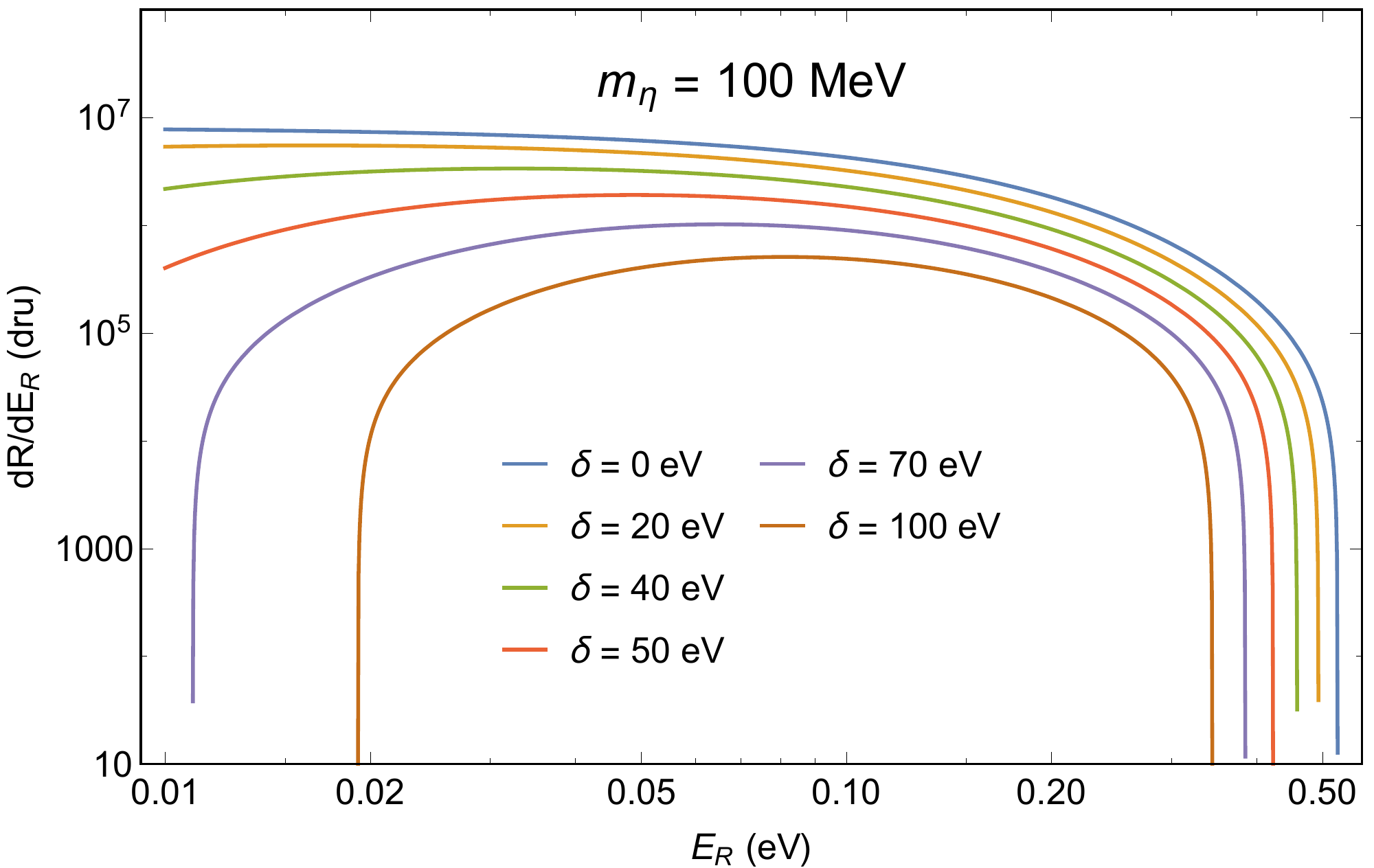}	
\captionsetup{labelfont = bf}
\caption{\label{fig:inelastic02}}
\end{subfigure}	
\captionsetup{justification   = RaggedRight,
             labelfont = bf}
\caption{\label{fig:inelastic}  Recoil spectrum   for inelastic scattering off a Xenon nucleus for different dark matter masses. We used, $m_{A^\prime}=$ 55 MeV and $V=$ 10 GeV. Note that  with the increasing values of $\delta$, the values of maximum recoil energy decrease. } 
\end{figure}

In Fig.~\ref{fig:crosssection}, we show both $\sigma_{SI}^{scalar(p,n)}$ and $\sigma_{SI}^{vector(p, n)}$. We have set $m_{\phi'} = 200$~MeV and $\delta=0$. Note that  the expression of $\sigma_{SI}^{vector(p, n)}$ does not have $m_{A'}$. They satisfy the before mentioned bounds. The dark matter-nucleus scattering cross section will be suppressed by a factor of $[1+(2m_A E_R)/m_{\phi',A'}^2) ]^{-2}$. Furthermore, the $A'$-mediated scattering is suppressed by an additional factor of  $[1-(2Z/A)]^2$. These factors will play important roles when we evaluate the nuclear recoil spectrum. We show the recoil spectrum for elastic process in Fig.~\ref{fig:elastic} and for inelastic process in Fig.~\ref{fig:inelastic} respectively. We have used Xenon ($A = $131 and $Z=$ 54) as the traget material and have expressed the differential event rate in ``differential rate unit" (dru) which is one event per keV per kg per day. Two interesting features of the recoil spectrum are, a) for the elastic case, the upper limit of recoil increases with the dark matter mass and b) for the inelastic case, the recoil becomes smaller with larger values of $\delta$ in order to satisfy the condition, $v_{min} \le v_{esc} $. 

  A variety of new techniques for probing low-mass dark matter are being studied, but few with the inelastico isospin-violating scattering in mind.  Since this is a generic phenomenon, it would be good to study these prospects (for some recent work, see~\cite{Bell:2021zkr,Bell:2021xff}).

\section{Relic density} \label{sec:relic density}

In this scenario, the correct thermal relic density of dark matter can be produced using the standard mechanism of freeze-out. The produced dark matter particles (co-)annihilate to either SM particles or to other dark sector particles. Several other non-standard mechanisms such as dark matter production from the decay of a heavy particle~\cite{Barrow:1982ei}, freeze-in mechanism~\cite{Hall:2009bx}, modifications to the expansion rate in the early Universe~\cite{Catena:2004ba} have also been used to obtain relic density of sub-GeV dark matter. It worthwhile to determine  if other mechanisms can be found for generating the correct relic density, which can expand the viable parameter  space. In our analysis we have assumed that $m_\eta > 40$~MeV to make sure that dark matter freezes out before BBN.

The most stringent  constraints on sub-GeV dark matter annihilation cross section comes from Planck data~\cite{Planck:2018vyg}, which constrains the effect of energy injection at the time of the recombination on the CMB. The Plank bound rules out the possibility of dark matter annihilating to SM particles through velocity independent $s-$wave channel, as the required cross section is very large. The following scenarios are consistent with the bounds, therefore they can be used to obtain the relic density,
\begin{itemize}
    \item {\it{First scenario}}. If the dark matter annihilation cross-section is $p$-wave suppressed i.e.  velocity dependent, $\braket{\sigma v} \propto v^2$, then it will be highly suppressed during the time of recombination as $v$ is very small at that time. Thus it can evade the Plank bounds.
    
     \item {\it{Second scenario}}. If two different species of dark matter co-annihilates at the time of freeze-out and the heavier component decays away completely by the time of recombination, then the co-annihilation cross section at the time of recombination will be negligible and satifies Plank bounds.
    
    \item {\it{Third scenario}}. If the final states produced in the dark matter annihilation processes are invisible then there is no extra energy injection during the recombination.
    
\end{itemize}

Two different channel are possible in our model for obtaining the correct relic density using the two light mediators.

\begin{itemize}

    \item $\phi'$-resonance: The dominant annihilation process can be the $s$-channel process mediated by the dark Higgs $\phi^'$, which is $p$-wave suppressed, with possible final states as $A' A', \bar \nu \nu, \bar \ell  \ell, \pi \pi, \gamma \gamma$, where the $\phi'$ is nearly on-shell. Note that, the $\phi^'$ couplings to the fermions are suppressed by the fermion masses and hence suppress the cross section. Therefore the process has to be at $\phi^'$ resonance in order to enhance the cross section. The annihilation cross section can be written as, 
    
    \bea												\sigma(\eta_i \eta_i \rightarrow \phi' \rightarrow X )v_{rel} &\sim&		\frac{ m_i^2 (E^2-m_i^2)}{4 V^2 E^2[(4E^2-m_{\phi^\prime}^2)^2+(m_{\phi^\prime} \Gamma_{\phi^\prime})^2]}				\times (2m_{\phi'}  \Gamma_{\phi'}) ,				\eea
    where $\Gamma _{\phi^'}$ is the total decay width of $\phi^\prime$.
    The correct relic density can be obtained using the resonance condition, 
    \begin{equation} 
    4 E^2 - m^2_{\phi^'} \gg \Gamma^2_{\phi^'} ,
    \end{equation}
    and in that case we need 
    \begin{equation}
         \frac{(4 E^2 - m^2_{\phi^'})^2/m^4_{\phi^'}}{\Gamma_{\phi'}/ m_{\phi^'}} \sim 10^4 . 
    \end{equation}

    \item $A'$-mediated: Co-annihilation of two different dark matter species mediated by $A^'$ is another possibility to generate the correct relic density in our model. The necessary conditions for this mechanism to work are that the mass eigenstates $\eta_1$ and $\eta_2$ have comparable abyndance during freeze-out, $\delta/m \le \mathcal{O}(0.1)$ and that the lifetime of $\eta_2$ is much greater than $\mathcal{O}(1)$~sec so that it decays away before recombination. The possible final states of the co-annihilation process are $\nu_A \nu_A$ and $e^+ e^-$, both of which are suppressed either by neutrino mixing angle or a kinetic mixing parameter. Note that the $\gamma \gamma $ final states is forbidden by Landau-Yang theorem~\cite{Yang:1950rg}. Therefore this mechanism does not play important role in our benchmark scenarios.
\end{itemize}

In Table.~\ref{tablerelic denitys }, we show two different benchmark scenario of the model parameters that can generate correct relic density of dark matter. Note that, both of them represent the $\phi^'$ resonance process. The corresponding dark  matter-nucleon cross section is also mentioned to show that they satisfy the direct detection constraints mentioned in Sec.~\ref{sec:direct detection}.

\begin{table}[h]

\captionsetup{justification   = RaggedRight,
             labelfont = bf}
\caption{ \label{tablerelic denitys } Two different benchmark scenarios have been considered and the corresponding values of the mass parameters and the cross sections are shown.}
            
\centering
\begin{tabular}{ llllllll }
\hline\hline
\makecell{$m_{A'}$  (MeV)}& $m_{\phi'}$ (MeV) & $m_\eta$ (MeV)&$m_{\nu_s}$(MeV)&$m_{\nu D}$(MeV)& \makecell{$\langle\sigma v\rangle$ \\ (cm$^3$/sec)}&											$\sigma_{SI}^{scalar}$(pb)&$\sigma_{SI}^{vector}$(pb)\\\hline	
&&&&&&& \\
\makecell{150}&\makecell{80}&\makecell{40}&\makecell{10}&\makecell{$10^{-3}$}&\makecell{3$\times 10^{-26}$}&\makecell{0.58}&\makecell{1.17}\\		\makecell{180}&\makecell{76}&\makecell{38}&\makecell{$10$}&\makecell{$10^{-3}$}&\makecell{3$\times 10^{-26}$}&\makecell{0.58}&\makecell{1.06}\\
\hline\hline
\end{tabular}
\end{table}

\section {Flavor Anomalies}  \label{sec:flavor physics}

In the flavor physics of B-meson there are various anomaly related to the observables coming from the process $b \rightarrow s \ell^+ \ell^-$. Two necessary conditions to address them are i) lepton flavor non-universality and ii) quark flavor violation. Our model has both of these ingredients to tackle the flavor anomaly problems.

\begin{itemize}
	\item the lepton flavor non-universality comes from the fact that the mediators $A^'$ and $\phi^'$ only couples to $\mu$ at tree level.
	\item flavor violating couplings in the quark sector comes from the UV-completion of the low energy model described in Sec.~\ref{sec:UV}
\end{itemize}

In the following, we restrict our analysis only to clean observables such as ~\cite{Geng:2021nhg} such as $R_K$, $R_{K^*}$, and $Br(B_s\rightarrow \mu^+\mu^-)$. The reason for such restriction is that these paramaters are devoid of hadronic uncertainties while other observables depend heavily on the hadronic form factor. The standard definition of $R_K$ and $R_{K^*}$ are, 

\bea
R_K &\equiv& \frac{Br(B\to K \mu^+\mu^-)}{Br(B\to K e^+e^-)} ~,
\\
R_{K^*} &\equiv& \frac{Br(B\to K^* \mu^+\mu^-)}{Br(B\to K^* e^+e^-)} ~.
\eea 

In SM, the lepton flavor couplings are universal and therefore the predictions for $R_K$ and $R_{K^{*}}$ are close to unity~\cite{Hiller:2003js,Bouchard:2013mia}. But the experimental results always contradict this prediction~\cite{RKstar,RKstarBelle, Aaij:2021vac,Aaij:2014ora,Aaij:2019wad}. The analysis of $R_K$ shows a 3.1$\sigma$ deviation from the SM prediction in the $q^2$ bin of 1.1 to 6 GeV$^2$ while for the $R_{K^*}$ the disagreement is at the 2.4$\sigma$ and 2.5$\sigma$ respectively for the $q^2$ bin of $(2 m_\mu)^2$ to 1.1 GeV$^2$ and 1.1 to 6 GeV$^2$ respectively~\cite{RKstar,RKstarBelle}. The latest results come from the analysis of data from RUN-1 and RUN-2 of the LHCb collaboration. In our analysis, we restrict ourselves to the central bin of $R_{K^*}$ data analysis due to the difficulty in explaining the data for both bin simultaneously using the effective operators. We consider another clean vobservable, the branching fraction of $b \rightarrow s \ell^+ \ell^-$ decay process. This was also reported by LHCb collaboration~\cite{LHCbseminar}. In the following we summarize the experimental results for all three observables.
\begin{eqnarray}
R_K &=& 0.846 ^{+0.042}_{-0.039}\textrm{(stat)}^{+0.013}_{-0.012}\textrm{(syst)},\\
R_{K^*} &=& 
\begin{cases}
0.660^{+0.11}_{-0.07}\pm 0.03  \,\, (2m_\mu)^2 < q^2 < 1.1~\mbox{GeV}^2 ~,
\\
0.685^{+0.11}_{-0.07}\pm 0.05 \,\, 1.1~\mbox{GeV}^2  < q^2 < 6~\mbox{GeV}^2 ~,
\end{cases} \\
Br(B_s\rightarrow \mu^+ \mu^-) &=& 3.09^{+0.46}_{-0.43}\textrm{(stat)}^{+0.15}_{-0.11}\textrm{(sysm)} 
    \times 10^{-9}.
\end{eqnarray}

\subsection{Theoretical Calculations}

The low energy effective theory of our model has lepton flavor non-universality and the UV complete high energy model generates flavor violating couplings in the quark sector. The $Z$ and $A'$ couplings to fermions in the flavor eigenstate basis are diagonal matrices which need not be proportional to the identity. But these couplings can be non-diagonal in the physical basis. This can lead to vertices of the form $\bar b \gamma^\mu P_{L,R} s (Z,A')_\mu$. The term $\bar b \gamma^\mu P_{L,R} s Z_\mu$  gives a contribution to $b \rightarrow s \ell^+ \ell^-$ process. On the otherhand, $\bar b \gamma^\mu P_{L,R} s A'_\mu$ contributes to $b \rightarrow s \mu^+ \mu^-$ process. Due to gauge invariance, such flavor changing vertices are not allowed for photon.

\begin{figure}[h]
\begin{subfigure}[h]{0.33\textwidth}
\includegraphics[width=0.8\linewidth,height=5cm]{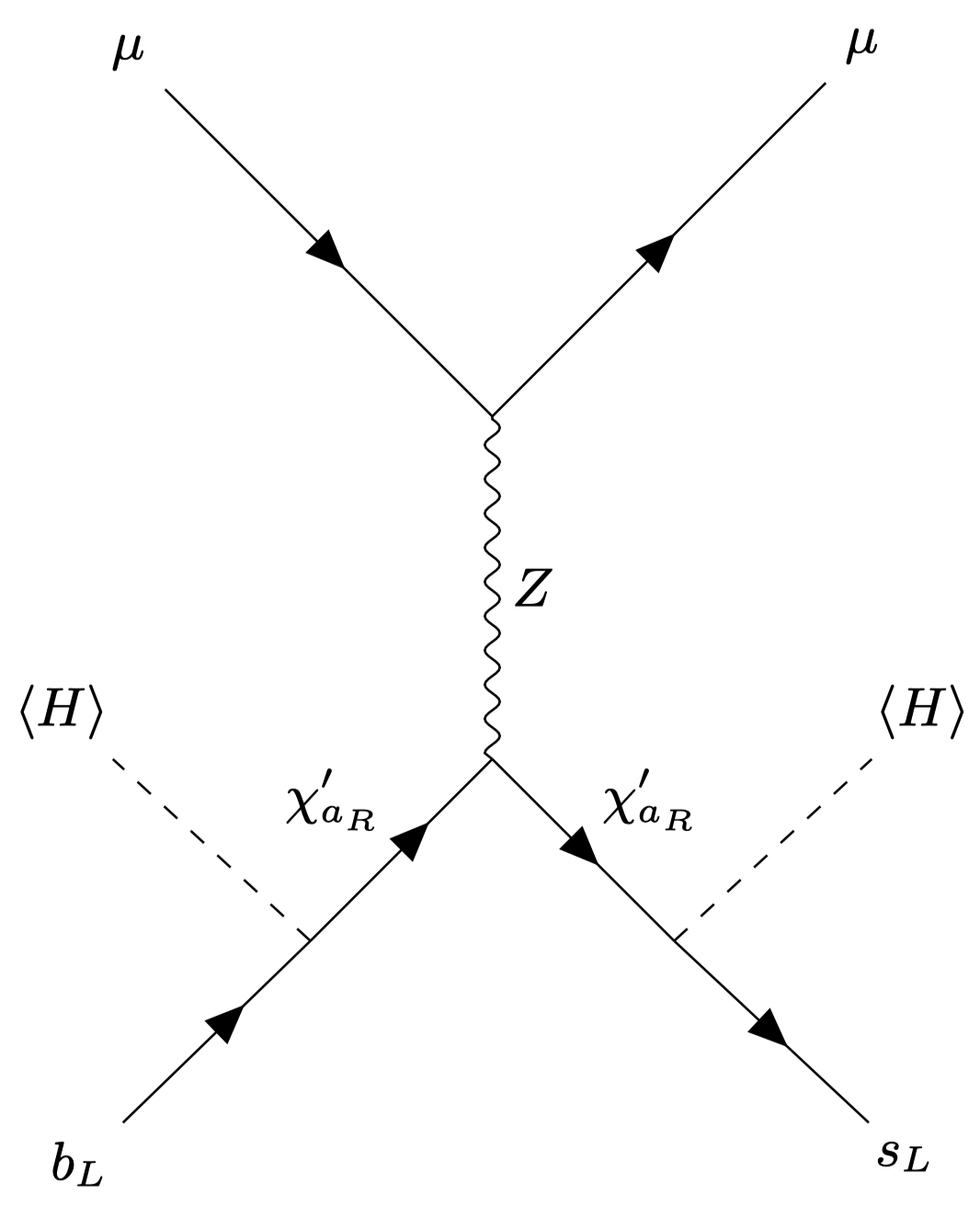}
\captionsetup{labelfont = bf}
\caption{\label{fig:RKU}}
\end{subfigure}	
\begin{subfigure}[h]{0.33\textwidth}
\includegraphics[width=0.8\linewidth,height=5cm]{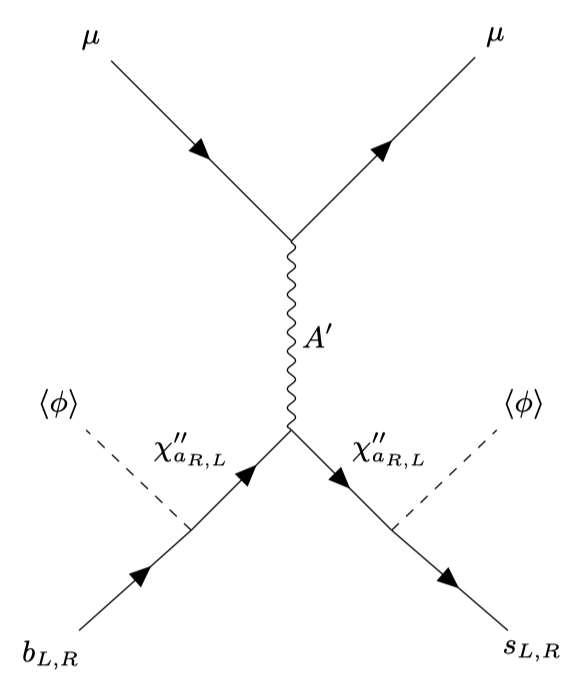}	
\captionsetup{labelfont = bf}
\caption{\label{fig:RKNU}}
\end{subfigure}	
\begin{subfigure}[h]{0.33\textwidth}
\includegraphics[width=0.8\linewidth,height=5cm]{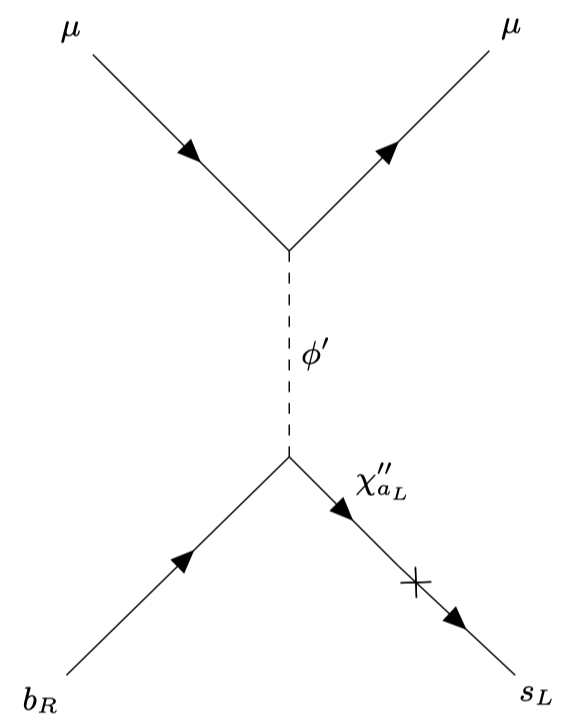}
\captionsetup{labelfont = bf}
\caption{\label{fig:RKscalar}}
\end{subfigure}
\captionsetup{justification   = RaggedRight,
             labelfont = bf}
\caption{\label{fig:RK} Feynman diagrams that contribute to the process $b \rightarrow s \ell^+ \ell^-$. } 
\end{figure}

Note, we have only considered a simple UV-completion in which we have added heavy fermions which couple to the fermions charged under $U(1)_{T3R}$.  But more generally, one could add more heavy fermions which couple to $b$ and $s$ in a similar manner, without generating anomalies. We consider an additional $\chi'_a$, neutral under $U(1)_{T3R}$, that mixes with $b$ and $s$ through  $\lambda'_{b,s} H \bar Q_L^{b,s} P_R \chi'_a + m'_{b,s}  \bar\chi'_a  P_R q_R^{b,s} + h.c.$ (assuming negligible mixing with the first generation). The $(\chi'_a)_R$ has same $Z$ coupling as $(b,s)_R$, therefore the $Z$-coupling to these right-handed quarks is the identity in every basis. But $(\chi'_a)_L$ has a $Z$ coupling which differs from $(b,s)_L$. This leads to a vertex of the form $\bar b \gamma^\mu P_L s Z_\mu$ at tree-level. This is shown in Fig.~\ref{fig:RKU}. Note that, a coupling of the form $\bar b \gamma^\mu P_L s A'_\mu$ is also induced at one-loop through $Z-A'$ kinetic mixing. But this term will be very small if we consider small kinetic mixing.

We add another vector-like fermion, $\chi''_a$. This is charged under $U(1)_{T3R}$  with charge $Q_{T3R} = 2$. The SM charges of this particle is same as $(b,s)_R$ and hence it can mix with $b,s$. The mixing is given by this term, $\lambda''_{b,s} \phi \bar \chi''_a P_R q_{b,s}$.  We assume negligible mixing with $d$. The interaction term $\bar b \gamma^\mu P_{L,R} s A'_\mu$ gets a tree-level contribution since $\chi''_a$ is charged under $U(1)_{T3R}$ while $b,s$ are not. This is shown in Fig.~\ref{fig:RKNU}. The coupling of $(\chi''_a)_L$ to $Z$ is different than $(b,s)_L$, therefore this term will give a tree-level contribution to the coupling $\bar b \gamma^\mu P_L s Z_\mu$. On the other hand,  there is no similar contribution to $\bar b \gamma^\mu P_R s Z_\mu$, since $(b,s$ and $\chi''_a)_R$ all have identical coupling to the $Z$ boson. Since $\chi''_a$ is charged under $U(1)_{T3R}$, a vertex of the form $\lambda''_{b,s} \phi' \bar q_{L(s,b)} q_{R(b,s)} \sin \theta'_{(s,b)L}$ is also possible. Fig.~\ref{fig:RKscalar} shows such possibilities.

The interactions that connect a ($b,s$) quark bilinear to a muon bilinear can be approximated with effective operators as for these processes the energy transfer is much larger than the mediator masses. In our scenario we get the following effective operators, 
\bea
{\mathcal{O}}_{U}^{Z} &=& 
\frac{e^2 }{3m_Z^2} 
\tan^2 \theta_W 
\left( \sin \theta_{sL} \sin \theta_{bL} + \sin \theta'_{sL} 
\sin \theta'_{bL} \right) \left(\bar b \gamma^\mu P_L s \right) \nonumber\\ &~& \times \left(\bar \mu \gamma_\mu 
\left[P_R + \left( 1-
\frac{1}{2 \sin^2 \theta_W} \right) P_L \right] 
\mu \right) \left(\bar b \gamma^5 s \right)(\bar \mu \gamma^5 \mu) \\ {\mathcal {O}}^{A'}_{NU} &=& \frac{1}{\Lambda^2} 
\sin \theta'_{s(L,R)} \sin \theta'_{b(L,R)} \left(\frac{m_{A'}}{\sqrt{2} V} \right)^2 
\left(\bar b \gamma^\mu P_{L,R} s \right)(\bar \mu \gamma_\mu P_R \mu) 
\nonumber\\&~& +
\frac{1}{\Lambda^2} 
\sin \theta'_{s(R)} \sin \theta'_{b(R)} \left(\frac{m_\mu m_b}{2 V^2} \right)\left(\bar b \gamma^5 s \right)(\bar \mu \gamma^5 \mu) 
\nonumber\\&~& -
\frac{1}{\Lambda^2} 
\sin \theta'_{s(L)} \sin \theta'_{b(L)} \left(\frac{m_\mu m_s}{2 V^2} \right) \left(\bar b \gamma^5 s \right)(\bar \mu \gamma^5 \mu)  \\ {\mathcal {O}}^{\phi'}_{NU} &=& 
\frac{\lambda''_s}{\Lambda^2} 
\sin \theta'_{bL} \frac{m_\mu}{\sqrt{2} V}
(\bar b P_R s) (\bar \mu \mu) 
 + \frac{\lambda''_b}{\Lambda^2} 
\sin \theta'_{sL} \frac{m_\mu}{\sqrt{2} V}
(\bar b P_L s) (\bar \mu \mu).
 \eea
 where the mixing angle $\theta_{(s,b)L}$ describes the mixing between the left-handed $(s,b)$ and $\chi'_a$; and $\theta'_{(s,b)(L,R)}$ are the 
left-/right-handed $(s,b)-\chi''_a$ mixing 
angles. Note that, we take $\Lambda \sim 2 $~GeV.

We use the following basis to expand the above operators, 
\bea
\frac{\alpha_{em} G_F}{\sqrt{2} \pi} V_{tb} V^*_{ts} \sum_{i,\ell} 
C_i^{bs\ell \ell} {\cal O}_i^{bs\ell \ell} ,
\eea
where different operators are given by,

\bea
{\mathcal {O}}_9^{bs\ell \ell} &=& (\bar s \gamma^\mu P_L b) (\bar \ell \gamma_\mu \ell),\hspace{2cm} {\mathcal {O}}_9^{'bs\ell \ell} = (\bar s \gamma^\mu P_R b) (\bar \ell \gamma_\mu \ell),
\nonumber\\
{\mathcal {O}}_{10}^{bs\ell \ell} &=& (\bar s \gamma^\mu P_L b) (\bar \ell \gamma_\mu \gamma^5 \ell) , \hspace{1.6cm} {\mathcal {O}}_{10}^{'bs\ell \ell} = (\bar s \gamma^\mu P_R b) (\bar \ell \gamma_\mu \gamma^5 \ell) ,
\nonumber\\
{\mathcal {O}}_{S}^{bs\ell \ell} &=& m_b (\bar s P_R b) 
(\bar \ell \ell) , \hspace{2.3cm}{\mathcal {O}}_{S}^{'bs\ell \ell} = m_b (\bar s P_L b) 
(\bar \ell \ell) ,
\nonumber\\
{\mathcal {O}}_{P}^{bs\ell \ell} &=& m_b (\bar s P_R b) 
(\bar \ell \gamma^5 \ell) ,
\hspace{1.9cm}
{\mathcal {O}}_{P}^{'bs\ell \ell} = m_b (\bar s P_L b) 
(\bar \ell \gamma^5 \ell).
\eea

Now we define, $C_i^U = C_i^{bsee}$ and 
$C_i^{NU} = C_i^{bs\mu\mu} - C_i^U$. Therefore the expansion of the operators give the following coefficients, 
\bea
\Delta C_{9}^{U} &=& (-146) (\sin \theta_{sL} \sin \theta_{bL} + \sin \theta'_{sL} 
\sin \theta'_{bL}) ,
\nonumber\\
\Delta C_{10}^{U} &=& (1.8 \times 10^3) 
(\sin \theta_{sL} \sin \theta_{bL} + \sin \theta'_{sL} 
\sin \theta'_{bL}) ,
\nonumber\\
\Delta C_9^{NU} &=& \Delta C_{10}^{NU} = 
(1.9 \times 10^8)
\sin \theta'_{sL} \sin \theta'_{bL} 
\left( \frac{m_{A'}}{\sqrt{2} V} \right)^2 ,
\nonumber\\
\Delta C_9^{'NU} &=&  
\Delta C_{10}^{'NU} =
(1.9 \times 10^8)
\sin \theta'_{sR} \sin \theta'_{bR} 
\left( \frac{ m_{A'}}{\sqrt{2} V} \right)^2 ,
\nonumber\\
\Delta C_{P}^{NU} &=&  
- \Delta C_{P}^{'NU} = -(2.0 \times 10^5 ~\mbox{GeV}^{-1})
\left(\frac{V}{10~\mbox{GeV}} \right)^{-2}
\nonumber\\
&\,& \times
\left(\sin \theta'_{sR} \sin \theta'_{bR} 
- (m_s/m_b) \sin \theta'_{sL} \sin \theta'_{bL} \right) , 
\nonumber\\
\Delta C_{S}^{NU} &=& (2.7\times 10^7 ~\mbox{GeV}^{-1})
\lambda''_b \sin \theta'_{sL} 
\frac{m_\mu}{m_b} \left(\frac{V}{10~\mbox{GeV}} \right)^{-1}, 
\nonumber\\
\Delta C_{S}^{'NU} &=& (2.7\times 10^7 ~\mbox{GeV}^{-1})
\lambda''_s \sin \theta'_{bL} 
\frac{m_\mu}{m_b} \left(\frac{V}{10~\mbox{GeV}} \right)^{-1} .
\nonumber\\
\eea

Note that, the universal lepton vector coupling is negligible as $\sin^2 \theta_W \sim 0.23$. The couplings and the mixing angles are free parameter, they can be tuned independently to control the coefficients.

\subsection{Numerical Calculations}

In general, it is usually very hard to address  the $R_{K^{(*)}}$ and $B_s\rightarrow \mu^+ \mu^-$ simultaneously in a new physics model with a vector mediator while satisfying  all current experimental bounds. The data from neutrino trident production at CCFR~\cite{Bauer:2018onh, Mishra:1991bv} and $Br(B\rightarrow K^{*}\nu\nu)$~\cite{Lutz:2013ftz}  tightly constraint the parameter space allowed by beam dump/fixed target experiments. One advantage of our model is that  the lack of the left handed neutrino couplings, because of which these bounds are not be applicable to our model. But on the other hand the chiral nature of the couplings give rise to another unavoidable  constraint in our model, which is $C_9^{(\prime)NU} = C_{10}^{(\prime)NU}$. Note that, the $R_K$ and $R_{K^{*}}$ measurements prefers a negative $C_9^{bs\mu\mu}$, or a positive $C_{10}^{bs\mu\mu}$ while the branching fraction of $B_s \rightarrow \mu^+\mu^-$ favors a positive $C_{10}^{bs\mu\mu}$, or a negative $C_{10}^{\prime bs\mu\mu}$.   Therefore an explanation of $R_K$ and $R_{K^{*}}$ with a  positive $C_{10}^{bs\mu\mu}$, which is favored by $B_s\rightarrow \mu^+\mu^-$, implies a negative $C_{9}^{bs\mu\mu}$. A negative $C_{9}^{bs\mu\mu}$ and a positive $C_{10}^{bs\mu\mu}$  imply a negative non-universal part and a positive universal part which means  a positive $C_{10}^{bsee}$ will leave the $R_K$ and $R_K^{*}$ unexplained. Therefore this 
constraint makes it difficult 
to explain the $R_K$ and $R_{K^{*}}$, and $Br(B_S\rightarrow \mu^+\mu^-)$ measurements simultaneously. But there are ways to overcome this difficulty. Specifically we consider following two scenarios.
\begin{itemize}
    \item {\it Scenario 1}: If we consider non-zero scalar and pseudoscalar couplings and use them to explain $Br(B_S\rightarrow \mu^+\mu^-)$. The $R_{K^{(*)}}$ can be explained by the other operators.
    \item {\it Scenario 2}: The other possibility is if we consider non-zero primed operators, which only contain the non-universal part. In that case the contributions are generated from both left-handed and right-handed quark couplings.
\end{itemize}

To do the numerical analysis we choose four different benchmark scenarios incorporating the above ideas. We calculate $R_K$ and $R_{K^{*}}$, and $Br(B_S\rightarrow \mu^+\mu^-)$ for all of them using flavio~\cite{Straub:2018kue}. We also calculate the SM pull, defined as $\sqrt{\Delta\chi^2}$ for each of them, considering only the clean observables from LHCb data. The SM pull helps us to understand how well those three measurements can be described and how significant the deviation is from the SM. In the following we describe the benchmark scenarios,
\begin{itemize}
    \item  The first three benchmark points correspond to the Scenario 1, where we introduce the scalar and/or pseudo-scalar operators. They were used to explain the $Br(B_s \rightarrow \mu^+\mu^-)$ while the others operators were used to fit $R_K$ and $R_{K^*}$. In particular, we introduce only scalar operators for BMA, while BMB and BMC have both scalar and pseudo-scalar operators. For BMA, $R_K$ and $B_s\rightarrow\mu^+\mu^-$ agree with the LHCb results within 1$\sigma$, while agreement of $R_{K^{*}}$  with the LHCb results is within 2$\sigma$ with a SM pull of 4.4$\sigma$. For BMB, all three observables agree with the LHCb measurements within 1$\sigma$, and the SM pull is 4.6$\sigma$. For  BMC, $R_K$ and $B_s\rightarrow\mu^+\mu^-$ agree  within 1$\sigma$, while $R_{K^{*}}$ is SM like with a SM pull of 3.8$\sigma$.  
    \item  For the fourth benchmark, BMD, which corresponds to Scenario 2, we introduce the primed operators that include non-universal parts. The agreement of $R_K$ and $B_s\rightarrow\mu^+\mu^-$  with the LHCb results is 1$\sigma$, and $R_{K^{*}}$ agree with the LHCb results within 2$\sigma$ with a SM pull of 4.2$\sigma$.
\end{itemize}

We summarize all the results in Table.~\ref{table:BM} along with the values coming from experimental data and the SM pull.

\begin{table}[h]

\captionsetup{justification   = RaggedRight,
             labelfont = bf}
             \caption{Explanation of $B$-anomalies using 4-benchmark points The first five rows 
           show the values of the coefficients 
             $C_{10}^{U}$, $C_{9,10}^{NU}$, 
             $|C_s - C'_s|$ (in units of $\gev^{-1}$), 
            $|C_p - C'_p|$ (in units of $\gev^{-1}$), 
            and $C^{'NU}_{9,10}$.  Rows 6-8 present 
            predictions for $R_K$, 
            $R_{K^*}$ (in the $q^2 \in [1.1,6] \gev^2$ 
            bin), and $Br(B_s \rightarrow \mu^+\mu^-)$.  Row 
            9 presents the SM pull of each benchmark point.
             \label{table:BM}}
            
\centering
\begin{tabular}{ c|c|c|c|c }
\hline\hline
&BMA&BMB &BMC&BMD \\
\hline
$C_{10}^{U}$ &4.85& -5.86&2.7 &-5.67\\
$C_{9,10}^{NU}$&-0.30 &3.65 &-0.8&4.55\\
$|C_s-C_s^{\prime}|$ GeV$^{-1}$& 0.033& 0.024 & 0.011& - \\
$|C_p-C_p^{\prime}|$ GeV$^{-1}$& - &0.030 &0.043 & - \\
$C_{9,10}^{\prime NU}$& -  & - &- &-1.28\\
\hline
$R_K$& 0.82 & 0.87&0.86 &0.87\\
$R_K^{*}[1.1,6]$ & 0.83&0.78 &0.97 &0.89 \\
$Br(B_s\rightarrow \mu^+\mu^-)$&3.36$\times 10^{-9} $ &3.05$\times 10^{-9}$ & 2.67$\times10^{-9}$&3.34$\times 10^{-9}$ \\
SM pull&4.4$\sigma$ &4.6$\sigma$ &3.8$\sigma$ & 4.2$\sigma$ \\\hline\hline
\end{tabular}
\end{table}

\section{Detection Prediction} \label{sec:detection}

In this section we discuss various possible detection prediction of our model in the future/upcoming low energy and high energy experiments.

\subsection{Various low energy neutrino experiments and FASER detection possibilities.}
 The new scalar and gauge boson can be produced from charged pion 3-body decays, i.e., $\pi \rightarrow \ell \nu_\ell A'(\phi')$. The three-body decay mode is not helicity-suppressed, unlike the two-body decay mode, and hence can be large. The couplings of $A'$ and $\phi$ must obey  experimental constraints on various charged pion decays. Recently, the three body decay of charged pion has been utilized to explain the MiniBooNE excess~\cite{Dutta:2021cip} which can potentially be accommodated in this model. The $A'$ also can be produced from the neutral meson decays. The FPFs  can explore both the light scalar and the gauge boson of his model.

We have mostly focused on the case in which the mediators decay visibly.  But FPF experiments an also probe the scenario in which the mediators decay invisibly through $A' \rightarrow \eta_1 \eta_2, \nu_R \nu_R$, $\phi' \rightarrow \eta_1 \eta_1, \nu_L \nu_R$.  In this case, the decay products may scatter against a distant target, producing events in FPF detectors such as FASER-$\nu$~\cite{FASER:2019dxq}.

\begin{figure}[h]
\centering
\includegraphics[height=10cm,width=14.2cm]{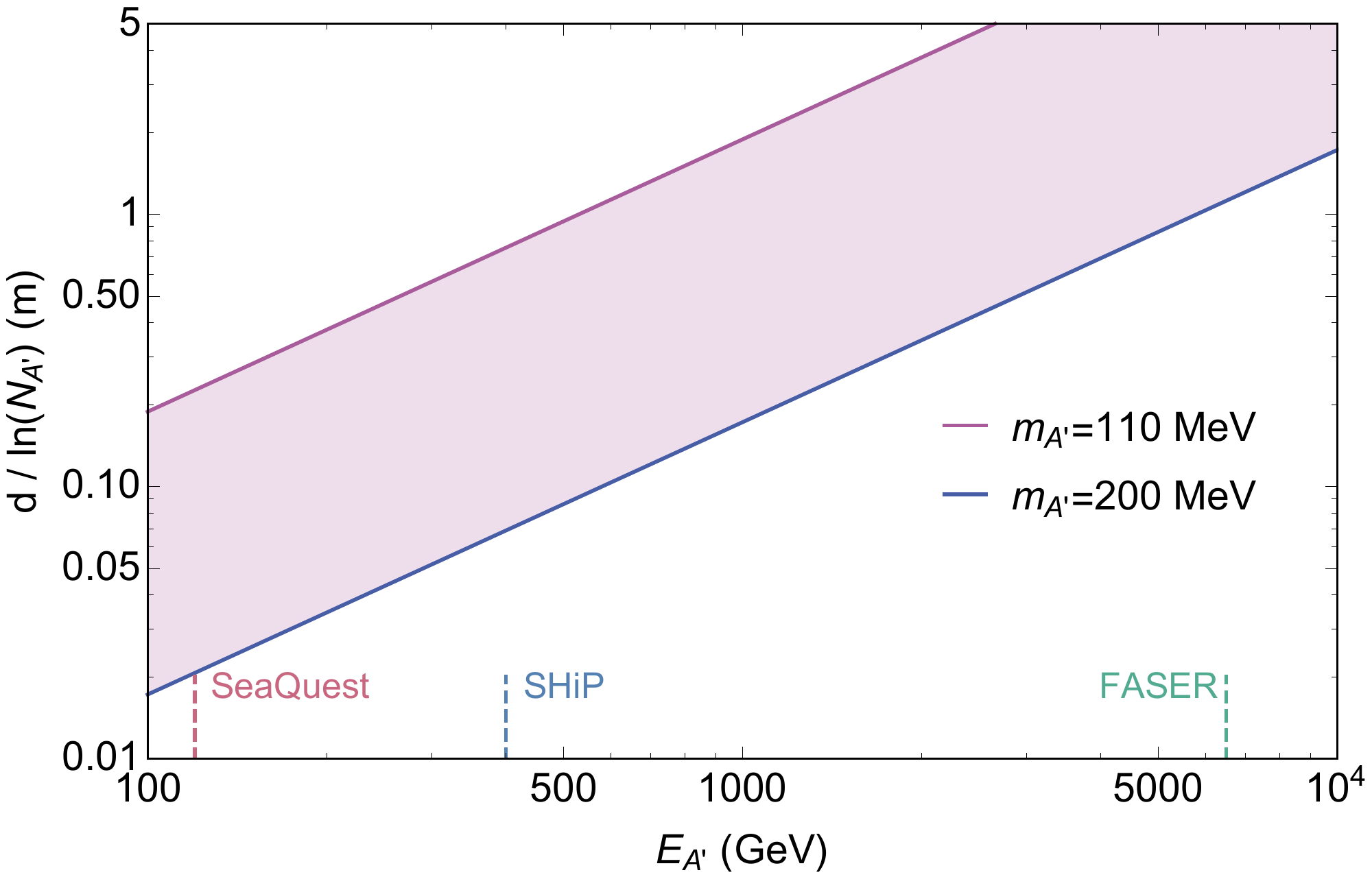}
\captionsetup{justification   = RaggedRight,
             labelfont = bf}
\caption{\label{fig:dvsEAplot} A rough estimate of maximum $d/\ln(N_{A'})$ necessary for an experiment to be able to probe this scenario for $m_{A'} \in [110\mev, 200\mev]$, as a function of the 
maximum $A'$ energy produced by the experiment~\cite{Dutta:2021afo}.  
$d$ is the displacement of the detector from the 
beam dump, and $N_{A'}$ is the number of $A'$ at 
energy $E_{A'}$ produced 
in a beam aimed at the detector.  The maximum 
$A'$ energies 
of FASER, SHiP and SeaQuest are also shown. }
\end{figure}

\subsection{LHC detection}

This scenario can be generalized to case in which $U(1)_{T3R}$ 
couples to top quarks.  In that case, the LHC can potentially 
discover new MeV-scale particles, such as $\phi'$, in conjunction 
with the heavy QCD-coupled particles.
We developed a unique  LHC search strategy for $\phi$ where it  is produced in association with a TeV-scale top-partner particle~\cite{Dutta:2022bfq}, denoted as $T$. 
The heavy top-partner particle is needed for UV completion as shown in Eqn. 5 which can be  produced at the LHC copiously via gluon interactions. $\phi$ will be produced from the decay of $T$ with substantial transverse momentum which causes the decay products of $\phi$ to be detected in the central region of the detector which will be energetic enough to overcome the SM backgrounds. This new search strategy will help LHC to utilized the reach of heavy top-partners to discover  MeV-scale $\phi$ which are  difficult to probe otherwise using traditional search strategies.
 \begin{figure}[h]
 \begin{center} 
 \includegraphics[height=11cm,width=15cm]{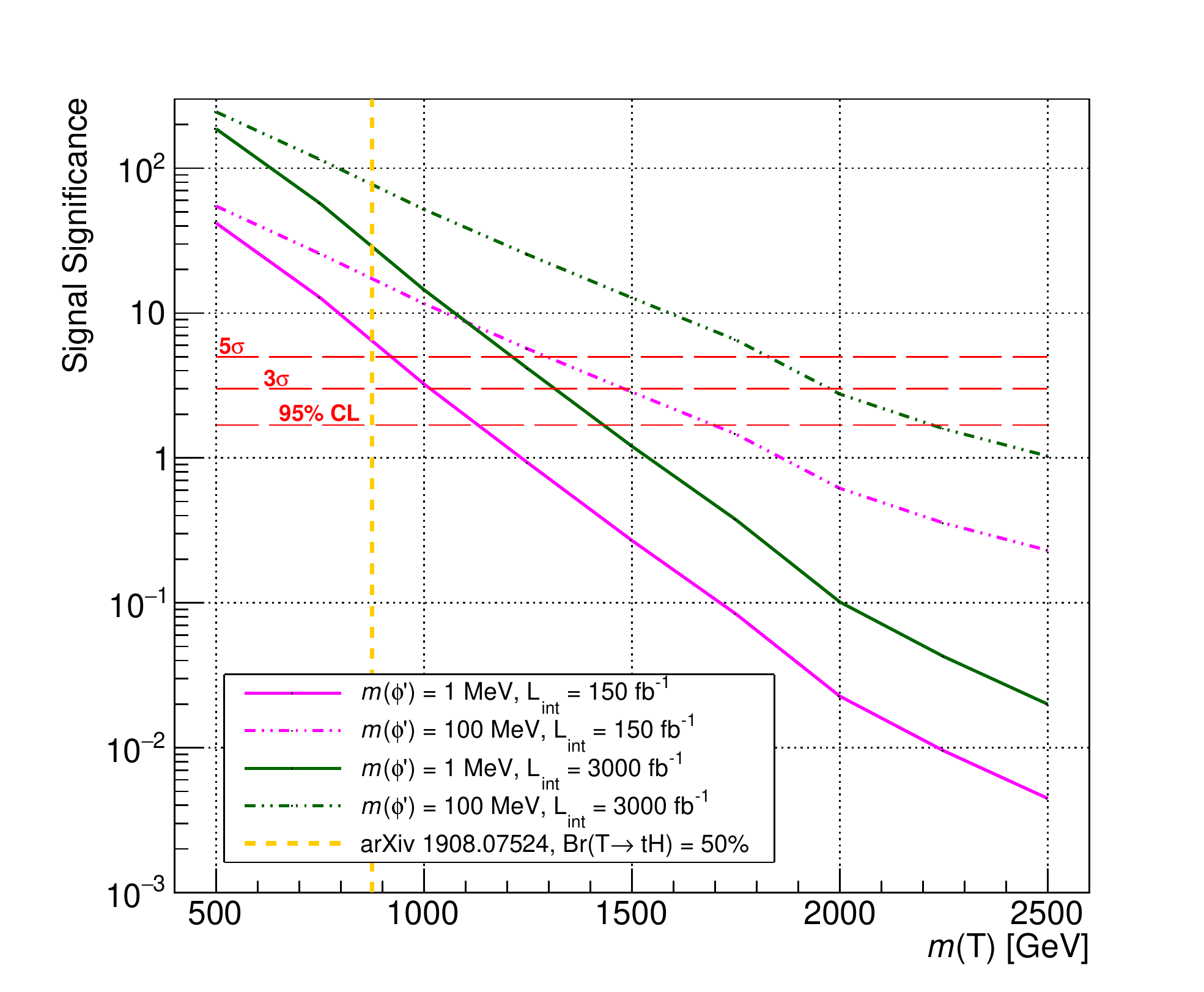}
 \end{center}
 \captionsetup{justification   = RaggedRight,
             labelfont = bf}
 \caption{Signal significance as a function of $m(T)$, for scenarios with $m(\phi')=1$ MeV (magenta) and  $m(\phi')=100$ MeV (green) and for scenarios with $\mathcal{L}_{int}=150\;\mathrm{fb}^{-1}$ (solid lines) and $\mathcal{L}_{int}=3000\;\mathrm{fb}^{-1}$ (dashed lines).   The dashed vertical yellow line indicates the lower limit on $m(T)$ found in~\cite{Cacciapaglia:2019zmj}, 
 assuming $\mathrm{Br}(T \rightarrow tH) = 50\%$.~\cite{Dutta:2022bfq}}
 \label{fig:Signal_Significance}
 \end{figure}
 Fig.~\ref{fig:Signal_Significance} shows the results of the expected signal significance for different  $m(T)$ and $m(\phi')$ scenarios. 
For the 150 fb$^{-1}$ scenario, it is feasible to exclude (at 95\% confidence level)  $ m(T) < 1.7$ $(1.1)$ TeV for $m(\phi') = 100$ $(1)$ MeV.

\section{Conclusion} \label{sec:conclusion}

In conclusion, we have constructed a sub-GeV dark matter model which utilizes an anomaly free $U(1)_{T3R}$ symmetry. The model is motivated to soften the hierarchy associated with the fermion masses from the first two generations. The masses of these fermions are associated with the symmetry breaking scale of $U(1)_{T3R}$ $\sim 1-10$ GeV which generates a sub-GeV scale for new vector and scalar mediators. The existence of two mediators help us to  satisfy  the thermal relic abundance. The parameter space of this model has constraints from various cosmological, astrophysics and low energy accelerator based constraints. In the allowed region of the  parameter space, the model can explain the observed anomaly of the $g-2$ of muon. The UV completion of the model allows it address the $R_{K^{(\ast)}}$ anomaly observed at the LHCb. The UV completion requires existence of heavy fermions in the theory which can be produced at the LHC and light mediators emerging from the decays of these heavy states can be also investigated at the LHC.

The light mediators of the models are connected to quarks which make the model visible at FASER and various beam-dump based neutrino experiments since these light mediators can be produced at these facilties from the neutral, charged mesons, bremsstrahlung and Drell-Yan process. The existence of focusing horn at some of these facilities will help the production from the charged modes. The model also should be able to address the MiniBooNE anomaly.

\newpage

{\bf Acknowledgments}

We would like to thank Peisi Huang, Teruki Kamon, Alfredo Gurrola and Dale Julson for their valuable contributions to the original papers based on which this white paper is written. The work of BD and SG are supported in part by the DOE Grant No. DE-SC0010813.  The work of SG is also supported in part by  National Research Foundation of Korea(NRF)’s grants, grants no. 6N021413. The work of JK is supported in part by DOE grant DE-SC0010504. We have used the package TikZ-Feynman~\cite{Ellis:2016jkw} to generate the Feynman diagram of Fig.~\ref{fig:loop}, and \ref{fig:RK}.

\bibliographystyle{utphys.bst}
\bibliography{ref.bib}

\begin{table}[p]
\captionsetup{justification   = RaggedRight,
             labelfont = bf}
\caption{ \label{table:result} A summary of the 
various experiments/probes considered here, their 
methods for producing and detecting the mediating particles, and the resulting sensitivities. }
\centering
\begin{adjustbox}{width=1.1\textwidth,center=\textwidth}
\begin{tabular}{ lllll }
\hline\hline

\makecell{Type of \\ experiments} & \makecell{Name of the \\ experiment} & \makecell{Production of $A^\prime/ \phi^\prime$} & \makecell{Final states} & \makecell{Results} \\
&&&& \\ \hline

\makecell{Electron \\  beam dump\\ experiments} & \makecell{E137, Orsay} & \makecell{$A^\prime$ : electron \\ bremsstrahlung\\ through kinetic \\  mixing at one-loop, \\ $\phi^\prime$ : Primakoff \\ production at \\ one-loop.} & \makecell{Both $A^\prime, \phi^\prime$ decay \\ predominantly \\ to visible SM 
\\ states $e^+e^-$. \\ $\phi^\prime$ decay is \\ rapid.} & \makecell{E137 rules out : \\ 1 MeV $\le m_{A^\prime} \le $ 20 MeV, \\ 1 MeV$\le m_{\phi^\prime} \le $ 65 MeV.\\ \\ Orsay rules out : \\ 1 MeV $\le m_{A^\prime} \le $ 40 MeV. \\} \\ 
&&&& \\

\makecell{Proton \\ beam dump \\ experiments}& \makecell{U70/NuCal, FASER \\ SHiP, SeaQuest \\ (displaced detector)} & \makecell{ $p$-bremsstrahlung \\ or meson decay \\ at tree level}& \makecell{$A^\prime \rightarrow e^+e^-$ \\ through kinetic \\ mixing. \\ $\phi^\prime \rightarrow  \gamma \gamma$ \\$\phi^\prime$ decays rapidly \\ hence cannot be probed. } & \makecell{ U70/NuCal rules out : \\ 1 MeV $\le m_{A^\prime} \le $ 93 MeV.\\  \\ FASER can probe : \\ 1 MeV $\le m_{A^\prime} \le $ 140 MeV. \\  \\ FASER 2/SHiP can probe : \\ 1 MeV $\le m_{A^\prime} \le $ 161 MeV. \\ \\  SeaQuest can probe : \\ 1 MeV $\le m_{A^\prime} \le $ 180 MeV. } \\
&&&& \\ 

\makecell{$e^+e^-$ collider \\ experiments}& \makecell{BaBar, Belle-II} &\makecell{ $e^+e^- \rightarrow \mu^+\mu^- + A^\prime/\phi^\prime $, \\ $e^+e^- \rightarrow \gamma A^\prime$}& \makecell{4$\mu$ final states, \\ $\gamma$ + invisible}& \makecell{BaBar rules out for \\ ($4\mu$ final states) : \\ 200 MeV $\le m_{A^\prime} \le $ 1.3 GeV, \\ 290 MeV $\le m_{\phi^\prime} \le $ 3 GeV. \\ \\  Belle-II can probe\\ ($\gamma$ + invisible): $m_{A^\prime} \ge 30$ MeV. } \\ 
&&&& \\ 

\makecell{$\bar{p} p$ collider \\ experiments} & \makecell{Crystal Barrel} &  \makecell{$ \bar{p} p \rightarrow \pi^0 \pi^0 \pi^0$, \\ $\pi^0 \rightarrow \gamma A'$} & \makecell{invisible states} & \makecell{ The parameter \\  space is ruled out for: \\ $55~ \text{MeV} < m_{A'} < 120~ \text{MeV}$ } \\
&&&& \\

\makecell{Fifth force \\ searches \\ experiments} & \makecell{Precision tests \\ of gravitational \\ Casimir, and \\ van der Waals forces}&\makecell{Relevant for extremely \\  light $A^\prime/\phi^\prime$. For $m_{A^\prime} \rightarrow 0$  \\ limit, the Longitudinal\\ mode  will contribute.}&\makecell{n/a}&\makecell{ The parameter \\ space is ruled out for: \\ $m_{A^\prime}/m_{\phi^\prime} \le 1$ eV. } \\
&&&& \\ 

\makecell{Astrophysical \\ probes} & \makecell{SN1987A, \\ Cooling of Sun \\ and globular clusters, \\ White dwarfs } &\makecell{ $\gamma + \mu \rightarrow A^\prime + \mu$,\\ $\mu + p \rightarrow \mu + p + A^\prime$,\\ $\mu^+\mu^- \rightarrow A^\prime$ at tree level,\\ $e^+e^- \rightarrow A^\prime$ through\\ kinetic mixing.}&\makecell{$A^\prime \rightarrow \eta \eta, \nu_s\nu_s$ (if decays to\\ $\nu \nu, e^+e^-$ then can not\\ escape),\\  $\phi^\prime \rightarrow \eta\eta,\nu \nu$ }&\makecell{SN1987A rules out : \\ $m_{A^\prime}, m_{\phi^\prime} \le 200$ MeV. \\ \\ Stellar cooling rules out:\\ $m_{A^\prime}, m_{\phi^\prime} \le 1$ MeV.\\  \\  WD constraint are negligible\\ if $m_\eta, m_{\nu_s} \ge 0.1$~MeV. \\  (All these astrophysical bounds can be\\ evaded  using chameleon effect.)} \\
&&&& \\ 

\makecell{Cosmological\\ probes} & \makecell{$\Delta N_{eff}$ value} & \makecell{ $\mu^+ \mu^- \rightarrow \gamma A^\prime$, \\ production of \\  longitudinal mode get\\ enhanced due to\\ axial vector coupling.}&\makecell{invisible states} & \makecell{If the Universe reheat at \\  a temperature $\ge 100$ MeV, \\ $m_{A^\prime}, m_{\phi^\prime} \le 1$ MeV is ruled out.\\ (Can be evaded if reheat occurs at \\a  lower temperature.)}\\ 
&&&& \\ 
\hline\hline
\end{tabular}
\end{adjustbox}
\end{table}

\begin{table}[t]
\centering
\begin{adjustbox}{width=1.1\textwidth,center=\textwidth}
\begin{tabular}{ lllll }
\hline\hline

\makecell{Type of \\ experiments} & \makecell{Name of the \\ experiment} & \makecell{Production of $A^\prime/ \phi^\prime$} & \makecell{Final states} & \makecell{Results} \\
&&&& \\ \hline

\makecell{Muon beam \\ experiments} & \makecell{NA64$\mu$, LDMX-M$^3$ \\ (nearby detectors)}& \makecell{$\mu-$bremsstrahlung}&\makecell{Can probe when \\ $A^\prime/\phi^\prime$ has a \\ significant decay rate\\ to invisible states \\ such as $\nu \nu, \eta \eta$}& \makecell{NA64$\mu$, LDMX-M$^3$ can probe \\ the entire parameter space \\ if $m_{A^\prime, \phi^\prime} > 2 m_{\eta,\nu_s}$ with \\ Br(invisible)$>10^{-4}$, \\ even if $A^\prime/\phi^\prime \rightarrow \mu^+\mu^-$ is allowed \\ still Br(invisible)$>10^{-4}$ \\ provided $m_{\eta,\nu_s} >1$ MeV. }\\ 

\makecell{Neutrino \\ 
experiments}& \makecell{COHERENT, CCM \\ JSNS$^2$} &\makecell{ $p/e$- bremsstrahlung, \\ meson decay }&\makecell{$A^\prime \rightarrow \nu_s \nu_s/\eta \eta $, \\ $\nu_s/\eta_i +N \rightarrow \nu_s/\eta_j +N$ \\generate nuclear recoil, \\ $\nu_s/\eta_i + e \rightarrow \nu_s/\eta_j + e$ \\generate electron recoil}&\makecell{Can be probed by looking at \\ nuclear/electron recoil. \\ \\ $m_{A^\prime} \sim 30$ MeV can explain the\\ 2.4-3$\sigma$ excess found by COHERENT, \\ $m_{A^\prime} \gtrsim 30$ MeV is ruled out.\\ \\  CCM and JSNS$^2$ will improve\\ the sensitivity. } \\
\hline\hline
\end{tabular}
\end{adjustbox}
\end{table}

\end{document}